\documentclass[twoside,preprint,amsmath,amssymb,floats,prfluids,lengthcheck,longbibiliography,onecolumn,11pt,tightenlines]{revtex4-2}
\usepackage[margin=1in]{geometry}
\usepackage{graphicx}
\usepackage[table, dvipsnames]{xcolor}
\usepackage{mathrsfs}[euscript]
\usepackage{enumitem}
\usepackage{eucal}
\usepackage{notes2bib}
\usepackage{booktabs}
\usepackage{boldline,multirow, hhline}
\usepackage{makecell}
\usepackage{array}
\usepackage{tabularx}
\usepackage{newtxtext,newtxmath}
\usepackage{titlesec}
\titlespacing*{\section}{0pt}{\baselineskip}{\baselineskip}
\titlespacing*{\subsection}{0pt}{\baselineskip}{\baselineskip}
\usepackage[bookmarks={false}, pdfstartview={XYZ null null 1.00}]{hyperref} 
\hypersetup{colorlinks  = true, urlcolor    = blue, linkcolor   = blue, citecolor   = blue,  filecolor    = blue}
\usepackage{fancyhdr}
\renewcommand{\footrulewidth}{0.5pt} 
\niibsetup{writekey=false,name={}}
\usepackage{filecontents}

\begin{document}
\count\footins = 1000


\title[Kulkarni et al.]{Increased solidification delays fragmentation \\ and suppresses rebound of impacting drops}
\author{Varun Kulkarni}
\altaffiliation[Also at: ]{School of Engineering and Applied Sciences, Wyss Institute for Biologically Inspired Engineering, Harvard University, Cambridge, MA 02138}
\author{Suhas Tamvada}
\altaffiliation[Also at: ]{Department of Mechanical and Aerospace Engineering, University of Florida, Gainesville, FL 32611}
\author{Nikhil Shirdade}
\author{Navid Saneie}
\altaffiliation[Also at: ]{Intel Corporation, Hillsboro, OR 97124}
\author{\\ Venkata Yashasvi Lolla}
\altaffiliation[Also at: ]{Department of Mechanical Engineering, Virginia Polytechnic Institute and State University, Blacksburg, VA 24060}
\author{Vijayprithiv Batheyrameshbapu}
\altaffiliation[Also at: ]{ANSYS Incorporated, Lebanon, NH 03766}
\author{Sushant Anand}
\email{Corresponding Author: sushant@uic.edu}
\affiliation{Department of Mechanical and Industrial Engineering \\ University of Illinois, Chicago, IL 60607}


\begin{abstract}
The splat formed after drop impact on supercooled solid surfaces sticks to it. On the contrary, a sublimating supercooled surface such as dry ice inhibits pinning and therefore efficiently rebounds drops made of a variety of liquids. While rebound is expected at lower impact velocities on dry ice, at higher impact velocities the drop fragments leaving behind a trail of smaller droplets. However, it is not known whether rebound can be entirely suppressed or fragmentation be controlled on such surfaces and if it depends on the extent of solidification inside the drop. In this work, we report on the role played by solidification within drops in modifying the outcomes of their impact on the supercooled ultra-low adhesive surface of sublimating dry ice. We show that the solidification thickness depends on the impact velocity and is the primary driver in suppression of rebound and also promotes a delay in fragmentation. Our findings imply that sublimating supercooled surfaces can present a broad spectrum of outcomes from complete bouncing to no-rebound which are not seen in drop impacts on supercooled superhydrophobic surfaces. We attribute this to thermo-elastocapillarity which considers bending of the solidified layer and is used to demarcate regime boundaries and determine the coefficient of restitution during rebound.\urlstyle{tt} \vspace{10pt}

\noindent \textit{This is not the latest version of the paper. The current version/preprint can be found at}: \\
{\small$\bullet$} \url{https://doi.org/10.1103/PhysRevFluids.9.053604} \textit{or}, \\
{\small$\bullet$} ResearchGate\textsuperscript{\textcopyright}, \textit{click} \href{https://www.researchgate.net/profile/Nikhil-Shirdade/publication/380560340_Increased_solidification_delays_fragmentation_and_suppresses_rebound_of_impacting_drops/links/664bf12022a7f16b4f3e534a/Increased-solidification-delays-fragmentation-and-suppresses-rebound-of-impacting-drops.pdf}{here} \textit{for pdf}.
\end{abstract}
\maketitle
\fancypagestyle{firstpage}
{
    \fancyhead[L]{KULKARNI \textit{et al}.}    
    \fancyhead[R]{2024}
    \fancyhead[C]{PHYSICAL REVIEW FLUIDS, Preprint, See abstract }
    \fancyfoot[R]{}
    \fancyfoot[L]{}
    \renewcommand{\footrulewidth}{0pt} 
}
\thispagestyle{firstpage}
\section{Introduction} \label{Sec1}
Liquid drop impact on supercooled surfaces maintained below the melting temperature of the solidified liquid ($T_m$) is decisive to several industrial applications and everyday life in icy climates \citep{Thievenaz2019, deRuiter2018, Han2012}. Depending upon liquid’s thermal and physical properties, substrate’s wettability and impact conditions, a fascinating spectrum of post-impact behavior is observed which ranges  from complete bouncing \citep{Yarin2006} in the absence of solidification to pinning-mediated adhesion of solidifying drops \citep{Seguy2023, deRuiter2018}. In the latter, adhesion of solidified material on surfaces, exemplified by ice accumulated on roads \citep{Riehm2012}, wind turbine blades \citep{Han2012}, and aircraft wings\citep{Lynch2001} is quite undesirable, disrupting everyday activities, industrial operations sometimes even imperiling human safety. Conversely, in applications like thermal spraying, splat quenching and additive manufacturing \citep{deRuiter2018, Thievenaz2019, Bhola1999} adhesion may be used advantageously by solidifying molten metals by impaction on an underlying substrate resulting in the formation of solid film of desired functional attributes. Whether the goal is to engineer surfaces and coatings for reduced ice adhesion or to create solid thin films with specific functional properties, achieving these objectives is crucially dependent on comprehending the outcomes of drop impact and the role of solidification during such impacts \citep{Schutzius2015, Kreder2016}. 

Fundamentally, the drop impact outcomes are dictated by liquid density $\rho_l$, surface tension, $\sigma_l$, and kinematic/geometric factors including impact velocity ($V_0$) and initial drop diameter ($D_0$) and succinctly described by the Weber number, \textit{We} which represents the ratio of the inertial ($\rho_lD_0^2V_0^2$) and surface tension force ($\sigma_lD_0$). Another factor besides these liquid properties which plays an important role is the nature of substrate as communicated by its wettability or roughness. Therefore, the liquid drop \textit{We} and the substrate properties, together can aptly describe both complete bounce (on highly non-wetting surfaces) - with/without fragmentation and, pinning-mediated adhesion \citep{Yarin2006}. 
\begin{figure}[t!]
\centering
\includegraphics[width= \textwidth]{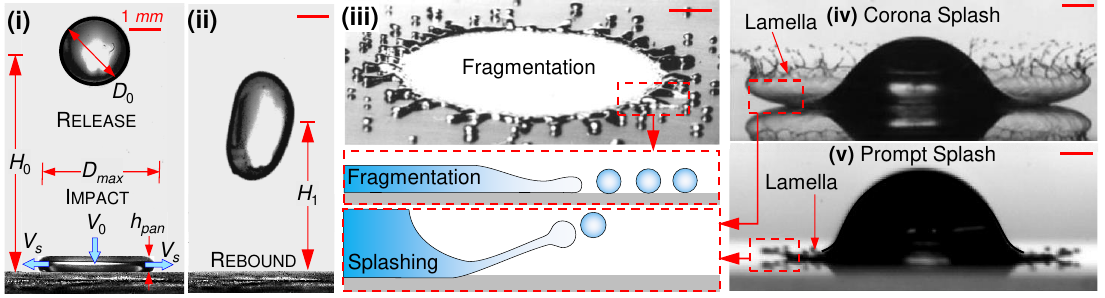}
\caption{\label{Fig1} \small Drop impact on a typical non-wetting surface, (\textit{i}) overlayed images of the release of a drop of diameter, $D_0$, from a height, $H_0$ and, moment of its impact with a velocity, $V_0$ forming a splat of maximum diameter, $D_{max}$, (\textit{ii}) rebound of a drop to height, $H_1$ (\textit{iii}) fragmentation of splat of molten tin (reprinted with permission, Aziz and Chandra \citep{Aziz2000}), (\textit{iv}) corona splash (reprinted with permission, Xu \citep{Xu2007}) (\textit{v}) prompt splash (reprinted with permission, Almohammadi and Amirfazli \citep{Almohammadi2019}). Dotted boxes in (\textit{iii}) - (\textit{v}) point to schematic side views of fragmentation and splashing. Scale bars from (\textit{i}) - (\textit{v}) represent a length of 1 mm.}
\vspace{-3mm}
\end{figure}

Physically, in simple terms we can understand rebound by considering a drop released from a height of $H_0$  (as shown in Fig. \ref{Fig1}\textcolor{blue}{(\textit{i})}, if its initial kinetic energy, $KE_0 \sim \rho_lD_0^3V_0^2$ exceeds its surface energy at maximum spread, $SE_{max}\sim \sigma_l D_{max}^2$ where, $D_{max}$ is the maximum horizontal splat diameter), i.e. $\Delta E = KE_0 - SE_{max} > 0$. The rebound height $H_1$ (as shown in Fig. \ref{Fig1} \textcolor{blue}{(\textit{ii})})  then can be simply estimated from $\Delta E$ by equating it to $\sqrt{mgH_1}$, where $m$ is the mass of the drop. From this, it may be tempting to infer that an increase in $KE_0$ with a comparatively lower increase in $SE_{max}$ leads to greater rebound (or, higher value of $H_1$), but, in practice, this increase is bounded. This is because, an increase in impact velocity, $V_0$ also increases the horizontal spreading velocity $V_s$  (as shown in Fig. \ref{Fig1} \textcolor{blue}{(\textit{i})},  whose spread is curtailed by surface tension. Consequently, liquid accumulates at the edge of the spreading drop forming a rim bounding a thinner central sheet of the denser fluid (drop), accelerating horizontally from 0 to $V_s = 3/8V_0$ into the lighter fluid (air) after the drop contacts the substrate \citep{Yonemoto2017}. Such spreading induces Rayleigh-Taylor instability \citep{Yarin2006}, with waves manifesting as fingers at the periphery of the rim that eventually disintegrate into smaller drops as shown in Fig. \ref{Fig1}\textcolor{blue}{(\textit{iii})} and the accompanying schematic. It is noteworthy that such splat fragmentation is differentiated from the frequently reported drop splashing \citep{Riboux2014, Garcia2021, Grivet2023}, which refers to the appearance of a lamella (liquid sheet) lifted upwards supported aerodynamically by a lift force\citep{Riboux2014} (shown schematically in the dashed box associated with Figs. \ref{Fig1}\textcolor{blue}{(\textit{iv}), (\textit{v})}. Usually, the expelled lamella in these cases either forms a bowl-like structure known as corona splash and shown in Fig. \ref{Fig1}\textcolor{blue}{(\textit{iv})} or an inclined sheet (this known as prompt splash as shown in Fig. \ref{Fig1}\textcolor{blue}{(\textit{v})} ultimately disintegrating into small droplets. 

A key factor in the dynamics described above are dissipative losses which slow the spread and recoil of the drop after impact. In isothermal impacts, it is commonly seen that the dissipation becomes significant when, (\textit{i}) dynamic viscosity of the liquid, $\mu_l$ is high and/or, (ii) there is an increase in viscous stresses ($\sim \mu_l V_0 h_{pan}$) due to reducing splat thickness, $h_{pan}$. The relative significance of dissipation due to viscosity can be assessed by two dimensionless groups: (i) Reynolds number, $Re$ which is the ratio of kinetic energy (inertia), $\rho_lV_0^2$ with the energy lost due to viscous dissipation, $\mu_l V_0D_0$  per unit volume resulting in, $\rho_lV_0D_0/\mu_l$ and, (ii) capillary number, $Ca$ given by the ratio of viscous dissipation to the surface energy, $\mu_lV_0/\sigma_l$. For instance, in liquids such as water, $Re \gtrapprox \CMcal{O}(10^{2})$ and $Ca \lessapprox \CMcal{O}(10^{-3})$ allowing us to ignore viscous effects, whereas for glycerol, $Re \lessapprox \CMcal{O}(1)$ and $Ca \gtrapprox \CMcal{O}(10)$ which indicates a non-negligible role of viscosity in drop dynamics.

Contrary to traditional, isothermal drop impacts on surfaces at room temperature described heretofore, impact of liquids (at room temperature) on supercooled surfaces presents non-isothermal conditions. Dissipation in such cases occurs thermally, through solidification, often represented in terms of the dimensionless Stefan number, $Ste = c_{p,s}\Delta T/ \CMcal{L}$ and Peclet number, $Pe=V_0D_0/\alpha_s$ where, $c_{p,s}$ (in kJ$\cdot$kg\textsuperscript{-1}K\textsuperscript{-1}) is the specific heat capacity of the solidified layer, $\CMcal{L}$ is the latent heat of solidification (in kJ$\cdot$kg\textsuperscript{-1}) and $\alpha_s$ (in m\textsuperscript{2}$\cdot$s\textsuperscript{-1}) is the thermal diffusivity of the solidified layer, respectively.  Solidification of the liquid drop here commences immediately upon contact with the surface and progresses gradually as the drop spreads enhancing pinning to the surface. Adhesion of such nature \citep{Seguy2023} influences the final splat morphology leading to intriguing outcomes such as self-peeling \citep{deRuiter2018} and fracture \citep{Ghabache2016}. Consequently, conditions leading to the arrest of drop spread are critical to the eventual shape assumed by these splats. In this regard, two main approaches have been adopted, the first only considers hydrodynamics at the drop scale and the second considers the dynamics at the contact line exclusively. In the first approach, at the drop scale level, solidification is either seen as reducing kinetic energy post impact \citep{Bhola1999} or cooling of the liquid is seen to augment dissipation which now consists of both the viscous and thermal boundary layer \citep{Thievenaz2020b}. The second approach focuses its attention to the contact line \citep{DeRuiter2017, Koldeweij2021} and argues that spreading of solidifying drops is arrested when at the contact line, (\textit{i}) the contact angle of the spreading drop equals the angle of the freezing front (\textit{ii}) a critical volume at the contact line is solidified or, (\textit{iii}) the liquid at the contact line reaches a critical temperature determined by the effect of kinetic undercooling \citep{Lolla2022}. Since the dynamics of drop spread are essential in realizing the eventual fate of the deforming drop, solidification can play a vital role in these impact scenarios.

The above survey highlights the role of adhesion and solidification in drop impact on supercooled surfaces and since they act in concert with each other, isolating their individual roles in the ensuing mechanics is often difficult.  For example, because drop pinning is omnipresent on engineered surfaces in non-isothermal impacts, the extent to which the solidification within a drop alone controls post-impact parameters such as maximal spreading, rebound height, contact time, rebound and splat-fragmentation is not clearly understood. The choice of test liquids and surfaces therefore becomes extremely critical. Even though water is one of the most preferred liquids for impact studies because of its large latent heat of fusion, surface tension and high supercooling, the effects of solidification within it during impact on supercooled non-wetting surfaces are small. In contrast, low surface tension liquids like alkanes have low supercooling, low dynamic viscosity and low heat of fusion making them prime candidates to study effects of solidification on drop impact. However, engineering surfaces that are completely non-wetting to them at all impact velocities is still challenging. One of the very few solid materials that successfully meets such stringent requirements is dry ice (\textit{di}), a supercooled (CO\textsubscript{2} gas) material that sublimates at $T_{di}$ = \textminus 78.9\textdegree C. Dry ice’s combination of being a molecular and sublimating solid eliminates any pinning between the drop and surface \citep{Antonini2013,Antonini2016} by providing near contactless levitation similar to Leidenfrost drops \citep{Tran2012} making it an ultra-low adhesive surface even for low surface tension liquids. Hence, we choose this material to isolate and show how solidification within a drop alone affects each of the above-mentioned facets of drop impact. While rebound and fragmentation of drops on dry ice has been known \citep{Antonini2013, Antonini2016}, we show that in the absence of pinning the extent of solidification controls both these outcomes delineating them from a region of no bounce when the splat merely spreads and rests on the surface. Since, the the layer closest to dry ice is solidified our examination concentrates on fragmentation rather than splashing which precludes any aerodynamic lifting of the lamella. Further, we also show that the rebound height in cases where rebound is seen is determined by the amount of solidification inside an impacting drop. In the course of our investigation, we also derive and use the dependence of solidification thickness on impact velocity which has not been tackled so far and provides a facile method to analyze such phenomena. Our efforts use a combination of laboratory experiments and theoretical arguments to explain the underlying physics behind our observations.

Following this introduction, which constitutes Section \ref{Sec1} of the paper, we organize the remaining text along the lines described here. We begin with a description of experimental details of the materials tested and experimental conditions in Section \ref{Sec2} following which we describe our experimental observations on the different drop impact morphologies in Section \ref{Sec3}. Thereafter, in Section \ref{Sec4} we provide the theoretical foundation for the dependence of solidification thickness on impact velocity extending \textit{Stefan}'\textit{s} analysis of gradual solidification of a liquid gently placed over a cold surface. Next, in Section \ref{Sec5} we describe the two observed regimes: fragmentation and rebound accompanied by scaling arguments to determine the criterion when these may be observed. In Section \ref{Sec6} we focus on the drop spread and derive the scaling for the maximum spreading of drop on impact. Finally, in Section \ref{Sec7} we discuss last of our results which details the effect of solidification on rebound height for different liquids. Our paper closes with Section \ref{Sec8}, which contains in brief, a summary of our findings, potential applications, open questions, and suggestions for future work.

\section{Methods, materials, liquid properties, impact conditions and dimensionless groups}\label{Sec2}
\vspace{-5pt}
\subsection{Setup and surface topography}
\begin{figure}[htp!]
\centering
\includegraphics[width=\textwidth]{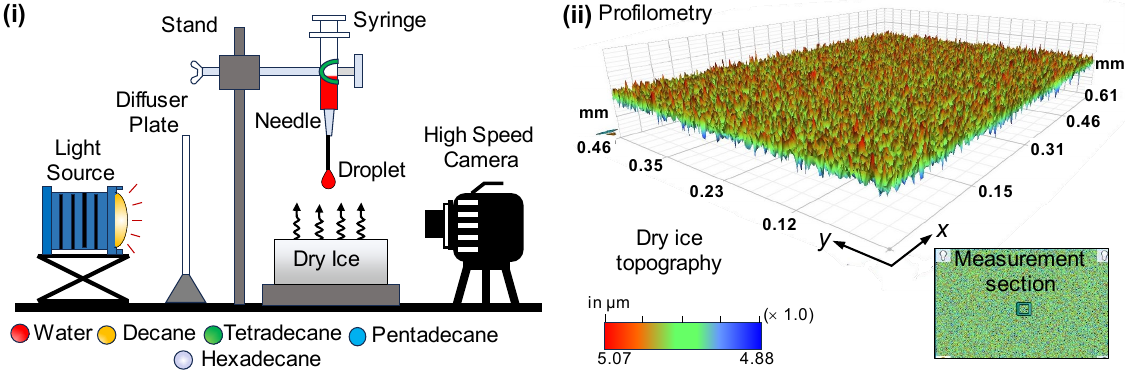}
\vspace{-3mm} 
\caption{\label{Fig2} \small (\textit{i}) Sketch of experimental setup showing impact of drop of on dry ice (not drawn to scale). Listed below are the tested liquids. (\textit{ii}) Topography of dry ice surface as measured using an optical profilometer giving an average value of roughness which equals 10 $\mu$m.}
\vspace{-3mm}
\end{figure}
Our setup (shown in Fig. \ref{Fig2}\textcolor{blue}{(\textit{i})}) consists of a needle attached to a glass syringe which is connected to a syringe pump and operated such that a single drop is ejected at a given time. Drops of diameters varying between 1.8$-$2.8 mm are tested and correspond to needles of gauge sizes 14, 16, 18, and 20 procured from Norsdon\textregistered. Five different test liquids, namely, water and long chain alkanes, decane, tetradecane, pentadecane and hexadecane are chosen as test liquids which shall be denoted by the color coding displayed below Fig. \ref{Fig2}\textcolor{blue}{(\textit{i})}. The capillary length, $\ell_{cap} =\sqrt{\sigma_l/\rho_lg}$ where $g$ is the acceleration due to gravity ($\approx 9.81$ m$\cdot$s\textsuperscript{-2}) for all these liquids is greater than or nearly equal to the drop diameters generated, ensuring that role of gravity is negligible in the drop deformation and impact dynamics. After their release from the needle the drops were allowed to impact on a dry ice slab at -78.9$^{\circ}$C. Room temperature/ambient conditions correspond to a temperature of 25$^{\circ}$C (or 298 K) and atmospheric pressure, 1 atm.

To observe and record the drop impact behavior we used Photron\textregistered FASTCAM Mini AX camera at 4000 frames per second (\textit{fps}) with a pixel a resolution of 1024 $\times$ 1024 pixels and a shutter exposure time of 5 $\mu$s. The selected frame rate gave us a temporal resolution of about 0.25 ms which was lesser than the impact time scale, $D_0/V_0$ of 1 ms. To spatially resolve the impact dynamics we used a high magnification lens  (InfiniProbe\textregistered TS$-$160) with a focal length between infinity and 18 mm which produced a magnification of 0$-$16$\times$ such that 1 pixel $\approx$ 15 $\mu$m. The background lighting used to illuminate our setup consisted of an LED (Nila-Zaila\textregistered) light source whose intensity of diffused and homogenized using an acrylic diffuser plates placed between the light source and the impacting drop. Drops of size, $D_0$ (in m) and density, $\rho_l$ (in kg$\cdot$m\textsuperscript{-3}) were released from heights, $H_0$ between 2 and 12 cm to vary their impact velocity ($V_0$) 0.3 and 1.5 m$\cdot$s\textsuperscript{-1} on nitrogen-purged frost-free dry ice surface. These experiment conditions correspond to Weber number, $We$ ranging from 12 to 120. The videos obtained were analyzed using the open source software IMAGE J \citep{Schneider2012}.

Lastly, the dry ice topography and roughness was experimentally measured using the Bruker-Nano Contour GT-K Optical Profilometer at the Nanotechnology Core Facility(NCF) at University of Illinois at Chicago(UIC). The samples were acquired with a 5 $\mu$m scanning step in both \textit{x} and \textit{y} perpendicular directions in the plane as shown in Fig. \ref{Fig2}\textcolor{blue}{(\textit{ii})} at a scanning velocity of approximately 2 mm$\cdot$s\textsuperscript{-1} in both directions. The three-dimensional topography revealed by these surface scans is shown in Fig. \ref{Fig2}\textcolor{blue}{(\textit{ii})} in the form of surface heights data acquired by the microprofilometer with a region of interest (ROI) within 1 mm\textsuperscript{2}. The non-uniformity of the dry ice surface is clearly seen with a height varying between  -5 $\mu$m and 5 $\mu$m. The experiments were repeated 5 times at different locations on the dry ice surface with a maximum standard deviation of  0.01 $\mu$m in the maximum and minimum height measured. Note that for a typical sublimation rate of 1$\%$ mass per hour, initial slab dimensions of $0.3\;\textrm{m}\;\times 0.15\;\textrm{m}\;\times 0.06\;\textrm{m}$, dry ice density of 1600 kg$\cdot$m\textsuperscript{-3} and experimentation time of 10 s the decrease in height of the sample is approximately 0.1 $\mu$m. This is $< 10\%$ of the measured roughness, $R_a$ $\approx$10 $\mu$m and therefore we can conclude that sublimation does not affect our profilometry measurements.
\vspace{0pt}
\subsection{Liquid thermal, physical properties, impact conditions and dimensionless groups}
For our experiments, we chose water and four alkanes: decane ($T_m \approx$ $- 30$\textdegree C), tetradecane ($T_m \approx$ 5\textdegree C), pentadecane ($T_m \approx$ 10\textdegree C) and hexadecane ($T_m \approx$ 18\textdegree C). These alkanes were chosen because except their $T_m$, their other thermal-fluid properties are all nearly the same so any difference in their post-impact behavior can be directly attributed solely to their thermal properties. 
\begin{table}[b!]
\caption{\label{tab:TableCombined} Thermal, fluid, interfacial properties, kinematic quantities and dimensionless groups}
\begin{center}
{  
  \begin{ruledtabular}
   \fontsize{8}{8}\selectfont
\begin{tabular*}{\textwidth}{l>{\centering\arraybackslash}m{40mm}>{\centering\arraybackslash}m{23mm}>{\centering\arraybackslash}m{12mm}>{\centering\arraybackslash}m{15.5mm}>{\centering\arraybackslash}m{15.5mm}>{\centering\arraybackslash}m{15.8mm}>{\centering\arraybackslash}m{15mm}}
		& & & \multirow{1}{*}{\textbf{Water}} & \multirow{1}{*}{\textbf{Decane}} & \multirow{1}{*}{\textbf{Tetradecane}} & \multirow{1}{*}{\textbf{Pentadecane}} & \multirow{1}{*}{\textbf{Hexadecane}} \\ \hline
	  \multirow{7}{*}{\shortstack[c]{Thermal\\ Properties}}	& \multirow{2}{*}{\textbf{Specific Heat} [kJ$\cdot$kg\textsuperscript{-1}K\textsuperscript{-1}] } & Liquid, $c_{p,l}$ & 4.18 & 2.21 & 2.19 & 2.20 & 2.22\\
&  & Solid, $c_{p,s}$ & 1.70 & 2.20 & 1.90 & 2.00 & 2.20\\ \cmidrule{2-3} 
                     & \multirow{2}{*}{\textbf{Thermal Conductivity} [W$\cdot$m\textsuperscript{-1}K\textsuperscript{-1}]} & Liquid, $k_l$ & 0.60 & 0.14 & 0.14 & 0.14 & 0.14\\
     &  & Solid, $k_s$ & 2.20 & 0.17 & 0.14 & 0.15 & 0.22\\ \cmidrule{2-3} 
   & \multirow{2}{*}{\textbf{Thermal Diffusivity} [m\textsuperscript{2}$\cdot$s\textsuperscript{-1}]}  & Liquid, $\alpha_l$ [$\times$ 10\textsuperscript{-7}] & 1.43 & 0.87 & 0.83 & 0.82 & 0.81\\
                     &   & Solid, $\alpha_s$ [$\times$ 10\textsuperscript{-7}]& 11.44 & 0.85 & 0.81 & 0.83 & 1.11\\ \cmidrule{2-3}
                     & \multirow{1}{*}{\textbf{Latent Heat} [{kJ$\cdot$kg\textsuperscript{-1}}]} & Liquid $\rightleftharpoons$ Solid, $\CMcal{L}$& 334 & 194 & 227 & 207& 236\\ \hline 
\multirow{4}{*}{\shortstack[c]{Fluid\\ Properties}} & \multirow{1}{*}{\textbf{Surface tension} [N$\cdot$m\textsuperscript{-1}]} & Liquid, $\sigma_l$ [$\times$ 10\textsuperscript{-3}] & 72 & 23.83 & 26.56 & 27.07 & 27.47\\ \cmidrule{2-3}
  & \multirow{1}{*}{\textbf{Dynamic viscosity} [Pa$\cdot$s\textsuperscript{-1}]}  & Liquid, $\mu_l$ [$\times$ 10\textsuperscript{-3}]& 0.89 & 1.26 & 2.33 & 3.10 & 3.30\\ \cmidrule{2-3}
  & \multirow{2}{*}{\textbf{Density} [{kg$\cdot$m\textsuperscript{-3}}]} & Liquid, $\rho_l$ & 998 & 728 & 764 & 769 & 770\\
                    &  & Solid, $\rho_s$ & 920 & 840 & 880 & 884 & 886\\ \hline  
  \multirow{2}{*}{\shortstack[c]{Impact\\ Conditions}} & \multirow{1}{*}{\textbf{Initial drop diameter} [m]} & \multirow{1}{*}{$D_0$ [$\times 10^{-3}$]} & 2.80 & 1.82& 1.88& 1.98& 1.98\\ \cmidrule{2-3}
      & \multirow{1}{*}{\textbf{Impact velocity} [m$\cdot$s\textsuperscript{-1}]} & \multirow{1}{*}{$V_0$} & \multirow{1}{*}{0.24$-$1.29} & \multirow{1}{*}{0.25$-$1.44} & \multirow{1}{*}{0.20$-$1.45} & \multirow{1}{*}{0.27$-$1.22} & \multirow{1}{*}{0.19$-$1.30} \\ \hline  
      \multirow{3}{*}{\shortstack[c]{Dimensionless\\ Groups }}& \multirow{1}{*}{\textbf{Stefan number}} &\multirow{1}{*}{\textit{Ste}}&\multirow{1}{*}{0.40} & \multirow{1}{*}{0.56} & \multirow{1}{*}{0.70} & \multirow{1}{*}{0.86} & \multirow{1}{*}{0.91}\\ \cmidrule{2-3}
        & \multirow{1}{*}{\textbf{Peclet number}}  & \multirow{1}{*}{{\textit{Pe}} [$\times$10\textsuperscript{4}]} &\multirow{1}{*}{0.53$-$2.84}&\multirow{1}{*}{0.52$-$2.98}&\multirow{1}{*}{0.48$-$3.54}&\multirow{1}{*}{0.65$-$2.90}&\multirow{1}{*}{0.59$-$3.12}\\ \cmidrule{2-3}
 & \multirow{1}{*}{\textbf{Weber number}} & \multirow{1}{*}{\textit{We} [$\times$10]} &\multirow{1}{*}{0.23$-$6.60}&\multirow{1}{*}{0.34$-$11.35}&\multirow{1}{*}{0.22$-$11.77}&\multirow{1}{*}{0.46$-$8.85}&\multirow{1}{*}{0.21$-$9.21}\\    
  \vspace{-4mm}                                                                       
		\end{tabular*}
  \end{ruledtabular}
}
\end{center}
\end{table}
The thermal, phase change and fluid properties for the liquid and solid phases of water and the four different alkanes tested are detailed in this section. These are used in this work and listed in Table \ref{tab:TableCombined} as: Thermal, fluid, kinematic and interfacial properties and dimensionless numbers. The elastic (Young's) modulus, $Y$ for solid alkanes is taken to be $2 \times 10^8$ Pa as alkanes fall under the category of paraffin wax while that for ice is $9.33 \times 10^9$ Pa \citep{Ghabache2016}. To calculate the thermal diffusivity for the solid ($\alpha_s$) and liquid ($\alpha_l$) phases we use the formula, $\alpha_s = k_s/\rho_s c_{p,s}$ and $\alpha_l = k_l/\rho_l c_{p,l}$. Also, the properties of the liquids in their solidified form are considered at their melting temperature, $T_m$ and their surface tension in their liquid form, $\sigma_l$ is considered to be within the same order of magnitude as that in solid form, $\sigma_s$. For water, $\sigma_s \approx 90$ mN$\cdot$m\textsuperscript{-1} and considering alkanes in their solid form to be similar to solid paraffin wax it amounts to approximately a value of 35 mN$\cdot$m\textsuperscript{-1} about 1.3 times their value in the liquid form. \textcolor{black}{Upon contact with dry ice substrate, the liquid immediately solidifies and therefore the interfacial tension between the solidified layer and dry ice is dropped from further analysis. The roles of other interfacial tensions that may arise are described in the relevant sections for regime transitions, Sections \ref{Sec5a} and \ref{Sec5b}}. The thermal conductivity of dry ice is, $k_{di} = 0.011$ W$\cdot$m\textsuperscript{-1}K\textsuperscript{-1}. All values presented here are taken from literature \citep{Yaws2006, Speight2017}.

A decrease in temperature when the drop touches the dry ice surface can lead to an increase in dynamic viscosity resulting in enhanced viscous dissipation, offering a plausible reason for the arrest of movement of the spreading drops. For this to be true, the viscous effects need to penetrate a larger thickness, typically the height of the pancake ($h_{pan}$) formed after drop impact, which would take longer compared to solidification of a thin micro-layer ($\delta$) of the drop in the vicinity of dry ice. \textcolor{black}{To emphasize the role of solidification over viscous effects, we can compare the time scales of spreading, $D_0/V_0$, viscous penetration, $\rho_l (D_0/4)^2/ \mu_l$, and thermal penetration ($\delta^2/\alpha_s$), using the length scales, $D_0/4$ (example, Fig \ref{Fig3} (iii), $t$ = 4 ms) and $\delta$ for viscous and thermal penetration respectively, at which arrest of spread may be expected. For typical conditions representative of our experiments, $D_0 = 2$ mm, $V_0 = 1$ m$\cdot$s\textsuperscript{-1}, $\alpha_s = 10^{-7}$ m\textsuperscript{2}$\cdot$s\textsuperscript{-1}, $\rho_l = 770$ kg$\cdot$m\textsuperscript{-3}, $\mu_l = $1 mPa$\cdot$s (from Table \ref{tab:TableCombined}) and $\delta \approx \CMcal{O}\left(10\right) \mu\textrm{m}$, we calculate the aforementioned times scales (approximately) as, 2, 125 and 1 ms respectively. This implies that only spreading and thermal penetration time scales are comparable, and hence viscous effects can be neglected.}

Similarly, for temperature dependent properties such as density, viscosity and surface tension to influence drop spreading, the thickness of the thermal boundary layer in time, $D_0/V_0$ corresponding to the drop impact/spreading time scale should be the order of the thickness of flattened drop. For our impact scenarios this value of the thermal boundary layer is $\CMcal{O} \left(10\right) \mu\textrm{m}$ m which means that the temperature decrease in the liquid drop is only restricted to a thin layer close to the sublimating dry ice surface. 

Hence, our choice of test liquids ensured that the role of temperature-dependent liquid properties or the role of viscosity was negligible. At room temperature, the non-dimensional Ohnesorge number, $Oh = \mu_l/\sqrt{\rho_l D_0 \sigma_l}$ for all liquids tested was relatively low assuming values between 0.0025 and 0.008. 

\section{Experimental Observations}\label{Sec3}
\begin{figure}[t!]
\centering
   \vspace{0cm}
    \includegraphics[width=\textwidth]{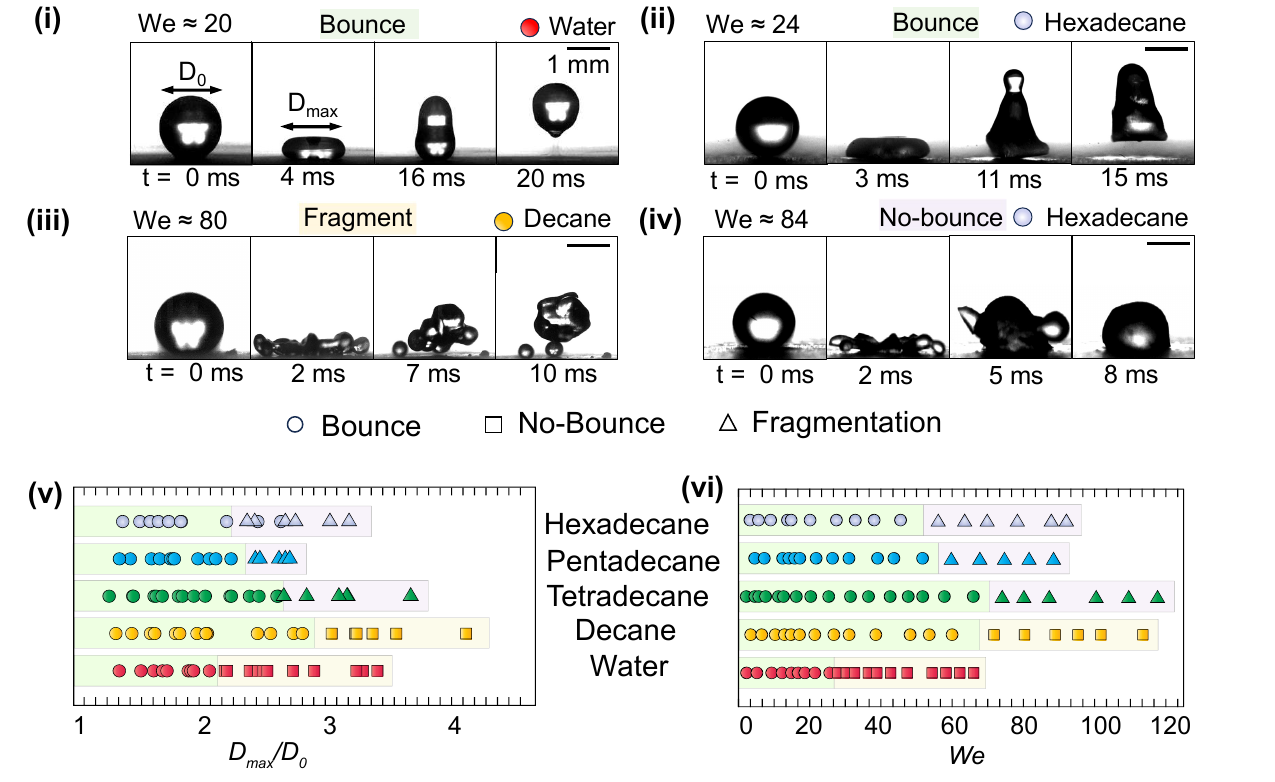}
   \vspace{-2mm} 
  \caption{\label{Fig3}Different regimes in drop impact (\textit{i}) bouncing with negligible solidification for water at \textit{We}$ \approx$ 20 (\textit{ii}) bouncing with discernible solidification for hexadecane, \textit{We}$ \approx$ 24 (\textit{iii}) fragmentation for \textit{We}$ \approx$ 80 (\textit{iv}) no-bounce for \textit{We}$ \approx$ 84. Raw data corresponding to transition regimes boundaries between bouncing/fragmentation and bouncing/no-bounce for different liquids as a function of (\textit{v}) $D_{max}/D_0$ and, (\textit{vi}) $We$.}
\end{figure}

Using the experimental methodology and liquids described above we systematically investigate two main aims: (\textit{i}) splat morphology after impact and its dependence on the impact \textit{We} and, (\textit{ii}) coefficient of restitution as measured by the rebound height. Our experimental findings addressing the first aim are displayed as a sequence of images in Fig. \ref{Fig3}\textcolor{blue}{{(\textit{i})} - {(\textit{iv})}}(also see SM \bibnote{See \href{SI_Kulkarni_et_al_2024.pdf}{Supplemental Material} for information related to the supplementary movies.} Movie 1, Movie 2). At low impact velocities all liquids bounce, similar to findings reported in literature \citep{Antonini2013, Antonini2016, Tran2012}. Here, to compare our results with these earlier studies we used water as the test liquid and impact it with dry ice surface at $We \approx 20$, an example of which is shown in Fig. \ref{Fig3}\textcolor{blue}{(\textit{i})}. We observe splat morphology and rebound behavior, identical to previous works\citep{Antonini2013, Antonini2016, Tran2012}. To examine the effects of solidification, we experiment with hexadecane which should exhibit higher solidification as it has higher \textit{Ste} (see Table \ref{tab:TableCombined}). For a hexadecane drop impacting at $We \approx 24$ shown in Fig. \ref{Fig3}\textcolor{blue}{(\textit{ii})} when solidification was substantial we continue to observe bouncing like water but with the major difference that the bottomost layer of drop looked partially solidified. Usually in supercooled surfaces solidification induced pinning restrict bounce off so this result on dry ice surface is unique especially since it proves that bouncing can be observed if adhesion is minimized or eliminated entirely.

To test whether bouncing continues or ceases we gradual increasing the higher impact velocities. At higher $We$ (or, impact velocities), two distinct outcomes emerge for almost the same $We$ (see SM \bibnotemark[1] Movie 3), the first being drop fragmentation (see Fig. \ref{Fig3}\textcolor{blue}{(\textit{iii})} for decane, $We \approx 80$) and the other wherein bouncing is suppressed altogether (Fig. \ref{Fig3}\textcolor{blue}{(\textit{iv})} for hexadecane, $We \approx 84$). This suggests that impact outcomes are dominantly dictated by solidification, which for low \textit{Ste} liquid like decane and water are shown to result in fragmentation producing a drop smaller in diameter which continues to bounce. In contrast, for liquids like hexadecane, pentandecane with high \textit{Ste} or increased solidification, bouncing is suppressed entirely when the thicknesses of the solidified layer increases. 

Lastly, we collate the raw data gathered from our experiments with all five liquids in terms of $We$ and $D_{max}/D_0$ and plot them as shown in Fig. \ref{Fig3}\textcolor{blue}{(\textit{v})} and \textcolor{blue}{(\textit{vi})}. Light green, light purple and light yellow color codes are used for demarcating regimes of bounce, fragmentation and no-bounce. It is to be noted that higher values of $We$ and $D_{max}/D_0$ typically imply higher impact velocity, $V_0$. 

\begin{figure}[t!]
   \centering
   \vspace{-2mm} 
    \includegraphics[scale=0.65]{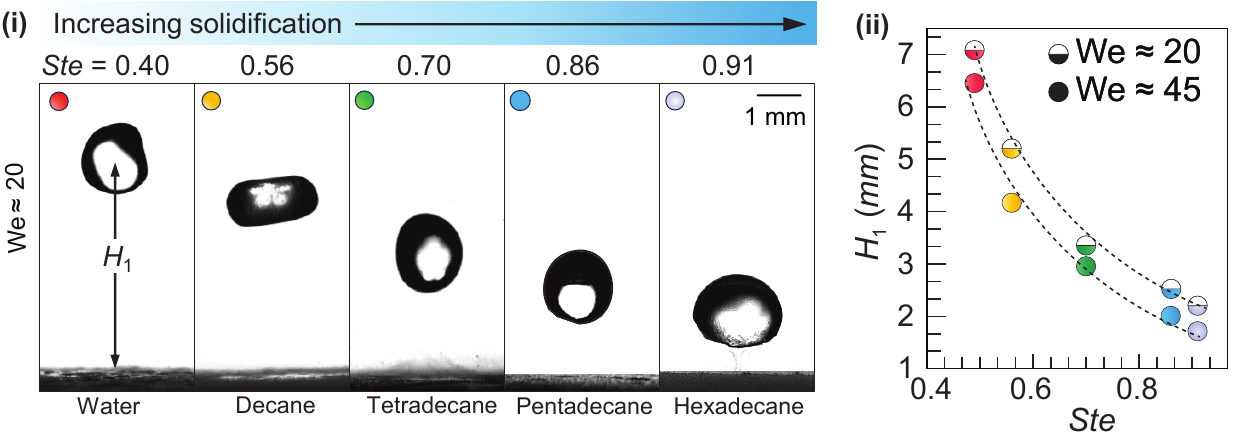}
   \vspace{-5mm} 
  \caption{\label{Fig4}(\textit{i}) Experimental images showing decrease in rebound height, $H_1$ with increasing solidification as indicated by the \textit{Ste} number. (\textit{ii}) $H_1$ at two different \textit{We} $\approx$ 20 (half filled semicircles) and 45 (filled semicircles), demonstrating two specific influences, first, the decrease in $H_1$ for all liquids with an increase in \textit{We} and, second, a decrease in $H_1$ at given \textit{We} for liquids with higher \textit{Ste} or with increased solidification.}
  \vspace{-1mm}
\end{figure}

In our second objective, we focus on the effect of solidification on the bounce height. All 4 alkanes and water are tested to understand this behavior in detail. A sample result from our experiments at $We \approx 20$ is shown in Fig. \ref{Fig4}\textcolor{blue}{(\textit{i})} (also see SM \bibnotemark[1] Movie 2). In this figure we see that from left to right, with increasing solidification, the rebound height decreases. Our results were further tested at different \textit{We} to observe whether for a particular liquid, an increase \textit{We} leads to a decrease in rebound height. Fig. \ref{Fig4}\textcolor{blue}{(\textit{ii})} depicts the raw data relating to two $We \approx$ 20 and 45 and we note that increasing $We$ leads to a decrease in rebound height for all liquids. Our measurements indicate that solidification not only depends on the liquid thermal and phase change properties but also impact velocity or \textit{We}. 
Both of above observations are studied in detail and rationalized theoretically in the next section (Sec. \ref{Sec4}) where effect of solidification with \textit{We} is quantified. We then use this information in deriving criteria for regime transitions in Sec. \ref{Sec5} followed by a understanding how drop spread is affected by solidification in Sec. \ref{Sec6} which an ancillary goal in understanding the drop dynamics of this nature better and conclude with Sec.\ref{Sec7} where our theoretical framework is used to explain our observation on rebound height.  

\section{Dependence of solidified liquid layer, $\delta$ on impact Weber number, $We$} \label{Sec4}

As suggested during the description of our experiments partial solidification is important in understanding the mechanics behind our observations. In this section we quantify it by calculating the thickness of the solidified portion of a liquid ($\delta$) when it comes in contact with a supercooled dry ice surface. An important consideration in this regard is that the surface of the dry ice is not smooth (see Section \ref{Sec2}) and the contact made by the drop is not perfect. Moreover, increasing $We$ can greatly affect this area of contact too, thereby increasing the extent of heat transfer. Additionally, the properties at the surface can differ significantly from that in the bulk. Together, these effects lead to development of non-negligible thermal contact resistance at this contact area developing a temperature distribution as shown in Fig. \ref{Fig5}\textcolor{blue}{(\textit{i})}. Two extreme limits of this at low and high \textit{We} with the difference in contact area are also sketched in Figs. \ref{Fig5}\textcolor{blue}{{(\textit{ii})}{(\textit{a})}} and \textcolor{blue}{{(\textit{iii})}{(\textit{a})}} with the resulting temperature distribution in Figs. \ref{Fig5}\textcolor{blue}{{(\textit{ii})}{(\textit{b})}} and \textcolor{blue}{{(\textit{iii})}{(\textit{b})}}. In the following we meticulously analyze cases high \textit{We} when contact resistance is negligible and then the case for arbitrary \textit{We} with finite contact resistance.

\subsection{Thickness of solidified liquid layer, $\delta$ at $We = 0$} \label{Sec1a}

\begin{figure}[b!]
   \centering
   \vspace{0pt}
    \includegraphics[width=0.8\textwidth]{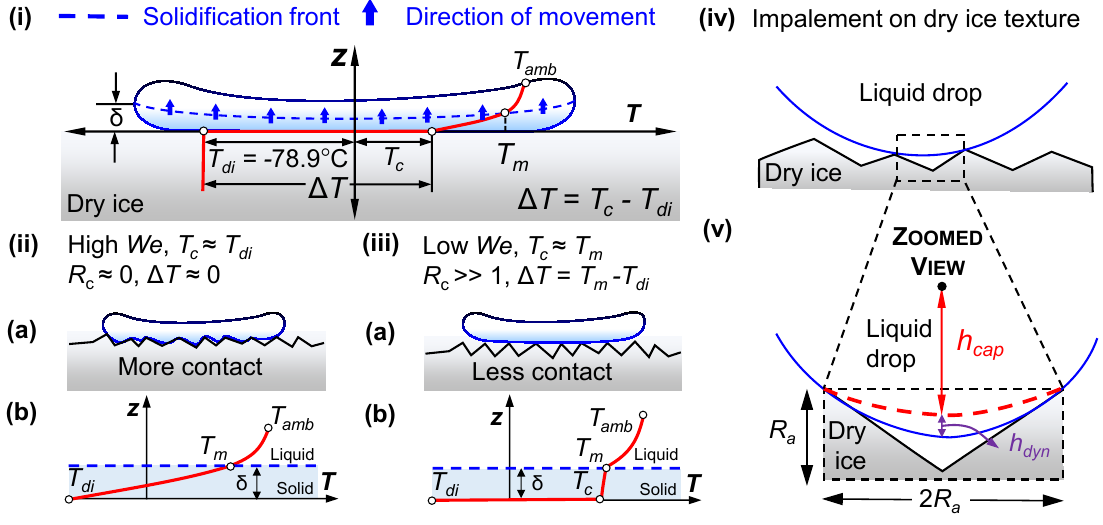}
   \vspace{-3mm} 
  \caption{\label{Fig5}(\textit{i}) General scenario of a drop in contact with a surface with non-negligible contact resistance, $R_c$ and the ensuing temperature distribution at the dry ice/liquid interface with a contact temperature of $T_c$. Portion of the solidified drop in contact with the supercooled surface is marked with a solidification thickness of $\delta$. Two limiting cases of contact resistance, $R_c$ when, (\textit{ii}) $R_c \approx 0$ when $We$ is high, sketched in (\textit{a}) with corresponding temperature profile in liquid and solid phases shown in (\textit{b}) depicting $T_c \approx T_{di}$. Note that red line within the zone of the solidified drop is a linear approximation to the unsteady temperature distribution which in the limit of small $\delta$ is an accurate representation of the temperature gradient at quasi steady or low \textit{Ste}. (\textit{iii}) $R_c >> 1$ when $We$ is low is sketched in (\textit{a}) with corresponding temperature profile in liquid and solid phases shown in (\textit{b}) depicting $T_c \approx T_{m}$ (\textit{iv}) Section of the drop in contact with the dry ice texture at one location  (\textit{v}) Exploded view showing depth of capillary penetration, $h_{cap}$ and dynamic pressure head due to drop impact, $h_{dyn}$ used to estimate contact resistance at a given \textit{We}.}
  \vspace{-2mm}
\end{figure}

We first consider the energy balance, commonly known as \textit{Stefan's condition} \citep{Thievenaz2019, Mills2016, Gielen2020} at the solidification interface \cite{Mills2016} and written as follows,
\begin{equation}\label{eqn:1a}
\rho_s  \CMcal{L} \dfrac{d\delta}{dt} = k_s{\dfrac{\partial T}{\partial z}}\biggr|_s - k_l{\dfrac{\partial T}{\partial z}}\biggr|_l
\end{equation}
Here, $\rho_s$ is the density of the solidified liquid, $\CMcal{L}$, is the latent heat of solidification of the liquid and $\delta$ is the thickness of the solidified layer. Eq. \ref{eqn:1a} is representative of the fact that the latent heat for solidification $\rho_s \CMcal{L} \frac{d\delta}{dt}$ is provided by the difference of conductive heat transfer in the liquid $-k_l{\frac{\partial T}{\partial z}}\bigr|_l$ and solid $-k_s{\frac{\partial T}{\partial z}}\bigr|_s$ phases across the solidification front (see Fig. \ref{Fig5}\textcolor{blue}{(\textit{i})}-\textcolor{blue}{(\textit{iii})}). $d\delta/dt$ is the velocity of the interface while it is solidifying, $k_s$ is the thermal conductivity of the solidified liquid and $k_l$ is the liquid thermal conductivity. In Eq. (\ref{eqn:1a}), $-k_l{\frac{\partial T}{\partial z}}\bigr|_l$ is much smaller compared to $-k_s{\frac{\partial T}{\partial z}}\bigr|_s$ as the thickness of the solidified layer, $\CMcal{O}(10^{-6})$ is much smaller than the diameter of the drop, $\CMcal{O}(10^{-3})$ and therefore the gradient, ${\frac{\partial T}{\partial z}}\bigr|_s$ is larger compared to ${\frac{\partial T}{\partial z}}\bigr|_l$. Here, $\Delta T_s = T_{m} - T_{di}$ (for solid phase) is $\CMcal{O}\left(10\right)$ and $\Delta T_l = T_{m} - T_{amb}$ (for liquid phase) is also $\CMcal{O}\left(10\right)$. Going forward, since $\Delta T_l$ will not be used, the subscript $l$ will be dropped and $\Delta T$ would mean $\Delta T_s$. 
Furthermore, we use the approximation $dT/dz$ as $\Delta T/\delta$ which is true for a quasi-steady approximation when $Ste < 1$ and implies that the temperature distribution within the solid layer is linear as shown in Fig. \ref{Fig5}\textcolor{blue}{(\textit{ii})}\textcolor{blue}{(\textit{b})} between the limits $T_m$ and $T_{di}$. We shall see in Section \ref{sec:prefactor} that this approximation is fairly accurate by considering the unsteady nature of the problem. Finally, using the foregoing simplifications we can express $\delta$ as,
\begin{equation}\label{eqn:3}
\rho_s  \CMcal{L} \dfrac{d\delta}{dt} = k_s \dfrac{\Delta T}{\delta}
\end{equation}
Integrating Eq. \ref{eqn:3} we derive the expression for $\delta$ using the definition of non-dimensional $Ste$, $\left( = \dfrac{c_{p,s} \Delta T}{ \CMcal{L}}\right)$ and the initial condition, $\delta \left(0\right) = 0$ as given below,
\begin{equation}\label{eqn:4}
\delta = \sqrt{2Ste\;\alpha_st} = 2\lambda\sqrt{\alpha_st}, \textrm{  where,  } \lambda = \sqrt{Ste/2}
\end{equation}
$Ste$ used here indicate fast solidification where large ($>1$) and small values ($<1$) correspond to slow solidification. Also, Eq. \ref{eqn:4} is very useful to determine the scaling for the solidification time scale by expressing $t$ as $t_{sol}$ and rearranging Eq. \ref{eqn:4}.
\begin{equation}\label{eqn:6S1}
t_{sol} \sim \dfrac{\delta^2}{\alpha_s Ste} = \dfrac{\rho_s \delta^2 \CMcal{L}}{k_s \Delta T}
\end{equation}
We shall use this expression as a time scale in the next section but before that we completely non-dimensionalize Eq. \ref{eqn:4} by evaluating Eq. \ref{eqn:3} at $t_{con} \approx 5D_0/V_0$ \citep{Bhola1999}, which denotes the time for which the drop is in contact with the dry ice surface. By defining, $\overline{\delta} = \delta/D_0$ we reduce Eq. \ref{eqn:4} to,
\begin{equation}\label{eqn:5}
\overline{\delta} = \sqrt{\dfrac{10Ste}{Pe}}
\end{equation}

The value of $\lambda$ in Eq. \ref{eqn:4} assumes a quasi steady approximation ($Ste < 1$) of the governing energy equation which gives rise to a linear temperature distribution within the solidified layer\citep{Ghabache2016}. For our case, \textit{Ste} though less than 1 is not significantly low to completely eliminate the time dependent term in the governing equation. We explore this in the next section and see if the quasi-steady (linear) approximation is actually reasonably justified for us.

\subsubsection{Determining the prefactor, $\lambda$}\label{sec:prefactor}
To calculate $\lambda$ corresponding to the exact non-linear temperature distribution in the solid and liquid layer we would need to solve the one dimensional heat equation \citep{Thievenaz2019} for both the phases with Stefan condition as one of the boundary conditions. This is expressed as below for the liquid (\textit{l}) and solid (\textit{s}) layers,
\begin{equation}\label{eqn:7S1}
\dfrac{\partial T_{l,s}}{\partial t} = \alpha_{l,s} \dfrac{\partial^2 T_{l,s}}{\partial z^2}
\end{equation}
At $z > \delta$, Eq. \ref{eqn:7S1} is solved for the liquid phase and for $z < \delta$ it is solved for the solid layer with the matching Stefan condition at the interface, $z = \delta$. A general solution \citep{Thievenaz2019, Mills2016} to Eq. \ref{eqn:7S1} can be written as,
\begin{equation}\label{eqn:7S2}
T_{l,s} = A_{l,s} + B_{l,s}\;\textrm{erf}\left(\dfrac{z}{2\sqrt{\alpha_{l,s}t}}\right) 
\end{equation}
The four constants $A_{l}, A_{s}, B_{l}$ and $B_{s}$ can be determined from the boundary conditions for the solid and liquid phases,
\setcounter{equation}{6}
\begin{subequations} \label{eqn:7S3}
\begin{align}
T_l &=  T_{amb} \quad\quad\quad\; \textrm{at} \;\;\; z \to \infty,\;\; t \geq 0   \;\;\;\;\;\; \textrm{(liquid layer)} \label{eqn:7a}\\ 
T_l &=  T_m  \quad\quad\quad\quad \textrm{at} \;\;\; z \to \delta,\;\;\;\; t > 0   \;\;\;\;\;\; \textrm{(liquid layer)} \label{eqn:7b}\\ 
T_s &=  T_{di} \quad\quad\quad\quad \textrm{at} \;\;\; z = 0,\;\;\;\;\; t \geq 0  \;\;\;\;\;\; \textrm{(solid layer)} \label{eqn:7c}\\ 
T_s &=  T_m  \quad\quad\quad\quad \textrm{at} \;\;\; z \to \delta,\;\;\;\; t > 0  \;\;\;\;\;\; \textrm{(solid layer)} \label{eqn:7d}
\end{align} 
\end{subequations}
The temperature distribution obtained in the solid and liquid layers using the boundary conditions above are therefore given by,
\begin{equation} \label{eqn:7S4}
T_s =  T_{di} + \dfrac{T_m - T_{di}}{\textrm{erf}\left(\lambda\right)}\;\textrm{erf} \left(\dfrac{z}{2\sqrt{\alpha_s t}}\right)
\end{equation}
\begin{equation} \label{eqn:7S5}
T_l =  T_{amb} - \dfrac{T_{amb} - T_m}{\textrm{erfc}\left(\lambda\sqrt{\alpha_s/\alpha_l}\right)}\;\textrm{erf} \left(\dfrac{z}{2\sqrt{\alpha_l t}}\right)
\end{equation}
In determining the temperature distributions in Eqs.\ref{eqn:7S4}, \ref{eqn:7S5} we have used the relations, $\textrm{erf}\left(q\right) = \int \displaylimits_0^q e^{-p^2} \, \mathrm{d} p$ and $\textrm{erfc}\left(q\right) = 1- \textrm{erf}\left(q\right)$ where, $\textrm{erf}\left(q\right)$ and $\textrm{erfc}\left(q\right)$ are the error and complementary error function, respectively. Furthermore, $\textrm{erf}\left(0\right) = 0$ and $\textrm{erf}\left(\infty\right) = 1$.
We also substitute for $\delta$ as $2\lambda \sqrt{\alpha_s t}$ in the expressions for the two temperature distributions above however it introduces another unknown $\lambda$. This can be evaluated using the Stefan boundary condition stated in Eq. \ref{eqn:1a} amounting to the transcendental Eq. \ref{eqn:7S6} which can be solved for $\lambda$.
\begin{equation}\label{eqn:7S6}
\sqrt{\pi}\lambda\;e^{\lambda^2}\textrm{erf}(\lambda) = Ste
\end{equation}
The roots of Eq.\ref{eqn:7S6} for different $Ste$ lead to $\lambda \approx \CMcal{O}(10^{-1})$ \citep{Mills2016}. For small $\lambda$ at small $Ste$, $e^{\lambda^2}\textrm{erf}(\lambda) \approx 2\lambda/\sqrt{\pi}$ leading to $\lambda = \sqrt{Ste/2}$ (derived in the next section). In this linear approximation, we see for our chosen values of \textit{Ste} (refer Table \ref{tab:TableCombined}) the values for $\lambda$ lie between $0.5-0.67$ and are in close agreement those obtained from the exact solution of Eq. \ref{eqn:7S6} which lie in the range, $0.45 - 0.6$.

\subsection{Thickness of solidified liquid layer, $\delta$ at any $We$} \label{Sec1b}

The expression for dimensionless solidified layer thickness, $\overline{\delta}$ ($=\delta/D_0$) at time, $t_{con} \approx 5D_0/V_0$ as given by Eq. \ref{eqn:5} predicts increasing values of $\delta/D_0$ at lower initial impact velocities, $V_0$ or $Pe$ (and by extension $We$). However, we expect the solidified layer thickness to decrease with lower $We$ as lesser contact is established at lower $We$ amounting to decreased heat transfer (see schematic Fig. \ref{Fig5}\textcolor{blue}{\textcolor{blue}{(\textit{iii})}}). To obtain this variation we recognize that contact resistance, $R_c$ can inhibit growth of the dimensionless solidified layer, $\overline{\delta}$. Here, we derive the relation for $\overline{\delta}$ with $We$ which includes the effect of contact resistance, $R_c$ to be used in theoretically determining the regime transition boundaries in Section \ref{Sec5} and COR calculations used in Section \ref{Sec7} pertaining exclusively to the regime where the drops bounce back. We divide our discussion into two subsections: first of which deals with determining $\delta =$ \textit{f}(\textit{t}, \textit{R}\textsubscript{c}), the second, on establishing the dependence of \textit{R}\textsubscript{c} with impact velocity($V_0$) or impact \textit{We}.

\subsubsection{Expression for $\delta$ as a function of contact resistance and time} \label{conressub1}

When a liquid drop touches the surface of dry ice heat is lost from the liquid drop to the composite dry ice/CO\textsubscript{2} thereby solidifying the drop. The CO\textsubscript{2} gas pockets lead to increase in contact resistance which becomes significant at lower impact $We$ (see Fig. \ref{Fig5} \textcolor{blue}{(\textit{iii})}). The transfer of heat occurs in three layers - (\textit{i}) contact layer comprising of the composite dry ice/CO\textsubscript{2} surface, (\textit{ii}) solidified liquid layer and, (\textit{iii}) the liquid layer (drop) as shown in Fig. \ref{Fig5}\textcolor{blue}{(\textit{i})}. Initially, the heat is transferred to the contact surface from the liquid drop which is at a higher (ambient) temperature. Thereafter, the liquid drop solidifies and creates another layer through which the heat is transferred via conduction to the remaining liquid drop. The heat transfer in these three cases may be written mathematically as follows \citep{Lipnicki2017},
\begin{enumerate}[label=(\itshape \roman*\textup),leftmargin=*,align=left]
\item Solidification of liquid drop by releasing latent heat, $\rho_s\CMcal{L}\dfrac{d\delta}{dt}$. 
\item Conduction through solidified layer to liquid drop, $k_s\dfrac{T_{amb} -T_c}{\delta}$. Here, $k_s$ is the thermal conductivity of the solidified liquid, $T_{amb}$ is the initial drop temperature and $T_c$ is the contact temperature of the interface when the drop meets dry ice. The linear temperature distribution within the solidified layer due to low $Ste$ justifies use of this simple expression (see section \ref{sec:prefactor} for further details).
\item Contact surface to solidified portion of the liquid drop, $\dfrac{T_c-T_{di}}{R_c}$
\end{enumerate}
Equating (\textit{i}), (\textit{ii}) and (\textit{iii}) above, we obtain,
\begin{equation}\label{eqn:8}
\rho_s\CMcal{L}\frac{d\delta}{dt}=k_s\frac{T_{m}-T_c}{\delta}=\frac{T_c-T_{di}}{R_c}
\end{equation}
The above three equations contain two unknowns, $\delta$ and $T_c$ which we can solve, bearing in mind that $R_c$ is known and shall be determined in section \ref{conressub2}. Considering last two of these expressions, (\textit{ii}) and (\textit{iii}), the contact temperature $T_c$ can be obtained in terms of $\delta$ as,
\begin{equation}\label{eqn:9}
T_c=\frac{T_{m}k_sR_c + T_{di}\delta}{k_sR_c + \delta}
\end{equation}
Exploring Eq. \ref{eqn:9} in the limits $R_c \approx 0$ and $R_c >> 1$ we obtain $T_c$ as $T_{di}$  and $T_m$ respectively. The temperature profile corresponding to these are sketched in Fig. \ref{Fig5}\textcolor{blue}{(\textit{ii})}\textcolor{blue}{(\textit{b})} and \textcolor{blue}{(\textit{iii})}\textcolor{blue}{(\textit{b})}. To solve for $\delta$ we consider the first two cases, (\textit{i}) solidification of liquid drop and, (\textit{ii}) conduction through solidified layer in Eq. \ref{eqn:8} and after a little algebra arrive at a non-dimensional form of the differential equation governing the evolution of $\overline{\delta}$ in time by choosing, $\overline{\delta} = \delta/D_0$ and $\tau = t/\left(D_0^2/\alpha_s Ste\right)$ where, $D_0^2/\alpha_s Ste = \tau_{sol}$ as defined in Eq. \ref{eqn:6S1} in Section \ref{Sec1a}. 
\begin{equation}\label{eqn:10}
\frac{d\overline{\delta}}{d\tau}=\frac{1}{(Bi^{cr}_s)^{-1}+\overline{\delta}}
\end{equation}
The Biot number of the solidified layer, $Bi^{cr}_s$ in Eq. \ref{eqn:10} is equivalent to the dimensionless contact resistance defined as, $D_0/k_sR_c$. Large values of $Bi^{cr}_s$ denote low contact resistance as usually assumed (see Fig. \ref{Fig5}\textcolor{blue}{(\textit{ii})}\textcolor{blue}{(\textit{a})} and \textcolor{blue}{(\textit{b})}) while finite, lower values signify increasing role of contact resistance (see Fig. \ref{Fig5}\textcolor{blue}{(\textit{iii})}\textcolor{blue}{(\textit{a})} and \textcolor{blue}{(\textit{b})}). With this physical description in mind we solve, Eq. \ref{eqn:10} subject to the initial condition, $\overline{\delta}\left(0\right) = 0 $ which represents the state when the solidified layer has not formed just when the drop touches the dry ice surface. The solution (neglecting the root which yields a negative value for $\overline{\delta}$) to Eq. \ref{eqn:10} is therefore given by,
\begin{equation}\label{eqn:11}
\overline{\delta}= \sqrt{(Bi^{cr}_s)^{-2} + 2\tau} - (Bi^{cr}_s)^{-1}
\end{equation}
In eqn \ref{eqn:11} the dependence of $Bi^{cr}_s$ on impact velocity or $We$ (in non-dimensional terms) remains to determined, which we shall pursue in the next section.
\subsubsection{Dimensionless contact resistance, $Bi^{cr}_s$ as a function of impact $We$}\label{conressub2}
One of the crucial considerations for estimating the contact resistance ($R_c$) is the area of the drop that is in contact with the composite dry ice/CO\textsubscript{2} gas surface. As one would expect, higher the impact velocity, more would be the area of contact compared to the case when the contact is purely due to gentle deposition. A simple way to account for this increase in contact area is by considering the ratio of depth the liquid penetrates due to impalement and total depth which includes the depth that is achieved upon gentle deposition \citep{Heichal2005}. The depth to which a liquid penetrates a texture ($h_{cap}$) can be calculated as, $h_{cap} = \sigma_l/\rho_lgR_a$ which is obtained by equating the Laplace pressure of meniscus of radius of curvature, $R_a$ and the hydrostatic pressure of column of height, $h_{cap}$ (see Fig. \ref{Fig5}\textcolor{blue}{(\textit{iv})} and \textcolor{blue}{(\textit{v})}). Whereas, the depth associated with impalement of the asperities is given by the Bernoulli dynamic pressure head, $h_{dyn} = V_0^2/2g$. The ratio, $f_s$ therefore assumes the form,
\begin{equation}\label{eqn:12f}
f_s  = \dfrac{V_0^2/2g}{\sigma_l/\rho_lgR_a + V_0^2/2g}
\end{equation}
On rearranging the above and defining, ${\overline{R}}_a = R_a/D_0$ we get,
\begin{equation}\label{eqn:12ffinal}
f_s  = \dfrac{We {\overline{R}}_a}{2 + We {\overline{R}}_a}
\end{equation}

\begin{figure}[htp!]
   \centering
   \vspace{0pt}
    \includegraphics[width=\textwidth]{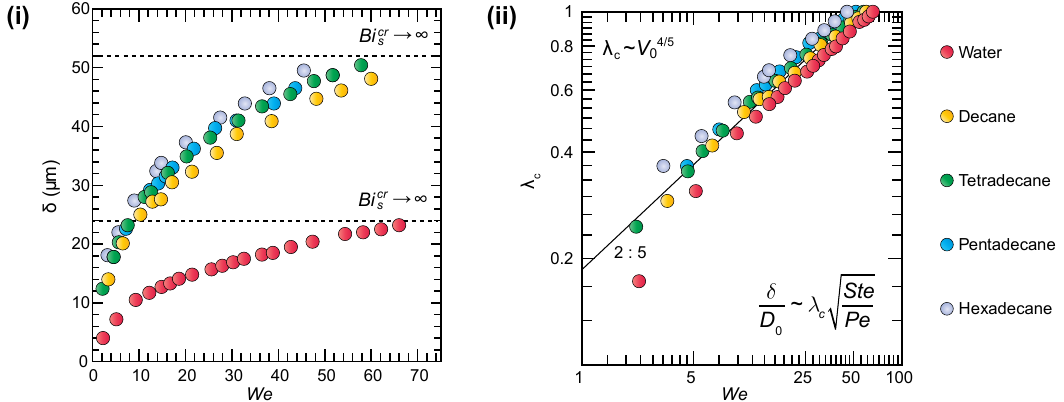}
   \vspace{-3mm} 
  \caption{\label{Fig6}(\textit{i}) Variation of dimensional solidified thickness $\delta$ with $We$ (in bouncing regime) as given by eqn \ref{eqn:12ext}. The dotted lines show the value of $\delta$ when contact resistance is negligible corresponding to $Bi^{cr}_s \to \infty$ (\textit{ii}) For any other $Bi^{cr}_s$ corresponding to a finite $We$, $\delta$ can be simply written as, $2\lambda\lambda_c \sqrt{\alpha_s t}$, where, $\lambda = \sqrt{Ste/2}$ and $\lambda_c = f \left(We \right) = 0.2We^{2/5}$ or $\sim V_0^{4/5}$. $\lambda_c$ assumes values from 0.2 to 1 which specifically for lower $We$ and $Bi^{cr}_s$ can be averaged to 0.5 and approximately 1 at higher $We$ or $Bi^{cr}_s$ as used in order of magnitude analysis for the scaling arguments to determine the regime boundaries (Fig. \ref{Fig8} \textcolor{blue}{(\textit{ii})} in Section \ref{Sec5}).}
  \vspace{0mm}
\end{figure}

In terms of heat transfer it means that the conduction across an interface with a roughness (depth) $R_a$, effective reference flat area, $f_s A_{\textrm{\textit{ref}}}$ (adjusted for the true contact area by the factor, $f_s$) and thermal conductivity of $k_{di}$ should be the same as that of an interface with contact resistance $R_c$ and area, $A_{\textrm{\textit{ref}}}$. For a temperature difference, $\Delta T$ across the interface such an energy balance implies, $k_{di} f_sA_{\textrm{\textit{ref}}} \Delta T/R_a = R_c^{-1} A_{\textrm{\textit{ref}}} \Delta T$ which yields, $f_s = R_a/R_c k_{di}$. Hence, $f_s$ can now be expressed as, ${\overline{R}}_a Bi^{cr}_{di}$. Eq. \ref{eqn:12ffinal} thus takes the form,
\begin{equation}\label{eqn:13ffinal}
Bi^{cr}_{di}  = \dfrac{We}{2 + We {\overline{R}}_a}
\end{equation}
Note that $Bi^{cr}_{di}$ and $Bi^{cr}_s$ are related to each other as, $Bi^{cr}_s = (k_{di}/k_s) Bi^{cr}_{di}$. Equating, $k_r = k_{di}/k_{s}$ and using Eq. \ref{eqn:13ffinal} in Eq. \ref{eqn:11} we obtain the dependence of $\delta$ on $We$ in the time the drop is in contact with the surface as,
\begin{equation}\label{eqn:12ext}
\delta = \sqrt{\left(\dfrac{2D_0 + We R_a}{k_r We}\right)^{2} + (2\alpha_sSte)\;t} - \left(\dfrac{2D_0 + We R_a}{k_r We}\right)
\end{equation}
For a roughness $R_a = 10\;\mu m$ (see Fig. \ref{Fig2}\textcolor{blue}{(\textit{ii})}) contact time, $t_{con} \approx 5D_0/V_0$ and $D_0$, $We$ and $Ste$ given in Section \ref{Sec2} we plot the variation of $\delta$ with $We$ obtained in expression \ref{eqn:12ext} as shown in Fig. \ref{Fig6}\textcolor{blue}{(\textit{i})}. Since, Eq. \ref{eqn:12ext} is cumbersome to use we see it equivalence with the commonly used form containing $\sqrt{t}$. Normally, it contains a prefactor $\lambda$ which physically represents the role of solidification in particular liquid (see Sec \ref{Sec1a}, Eq. \ref{eqn:4}) when contact resistance effects are not included. For finite $Bi^{cr}_s$ we consider a form analogous to the traditional form but includes a multiplicative factor $\lambda_c$ to account for the difference caused due to a non-negligible contact resistance,
\begin{equation}\label{eqn:delfin}
\delta = 2\lambda_c\lambda \sqrt{\alpha_s t} \textrm{  with  } \lambda = \sqrt{Ste/2} \textrm{  and  } \lambda_c = f(Bi^{cr}_s) \textrm{  or  } f(We)
\end{equation}
Our results for $\delta$ using Eq. \ref{eqn:12ext} are now plotted as function of $We$ in Fig. \ref{Fig6}\textcolor{blue}{(\textit{i})}. The curve so obtained can now be compared with Eq. \ref{eqn:delfin} to determine $\lambda_c$, our correction to the standard expression which does not include effects of contact resistance. We notice from Fig. \ref{Fig6}\textcolor{blue}{(\textit{i})} that $\delta$ decreases by a factor 5 at very low \textit{We} or $Bi^{cr}_s$ such that $\lambda_c$ is $\CMcal{O}(10^{-1})$ at lower $We$ with an average value of 0.5 at these lower values and close to 1 at higher $We$ or $Bi^{cr}_s$. We specifically plot $\lambda_c$ in Fig. \ref{Fig6}\textcolor{blue}{(\textit{ii})} and see that it yields a dependence of the form, $\lambda_c \sim We^{2/5}$ or $V_0^{4/5}$.  As $Bi^{cr}_s \to \infty$, (representing no contact resistance) we recover the standard result, $\delta = \sqrt{2Ste \alpha_s\;t}$ which is shown by dotted lines in Fig. \ref{Fig6}\textcolor{blue}{(\textit{i})} and implies $\lambda_c \to 1$. 
We observe that water has lower values for $\delta$ compared to the alkanes. This is expected since it has the lowest $Ste$ which implies slow solidification and thereby its effective thermal diffusivity ($\alpha_{\textrm{eff}} = \alpha_s Ste^2$) \citep{Thievenaz2019, Thievenaz2020a} is the least, differing by almost factor of 4 from water to the alkanes which is seen as a difference in solidification thickness by a factor of 2 at high $Bi^{cr}_s$ or $We$.

\section{Regime boundaries}\label{Sec5}
\subsection{Transition boundary between fragmentation and no-bounce}\label{Sec5a}
At large \textit{We}, a (flattened) drop close to its maximum spread ($D_{max}$ at time, $t = t_{max}$) displays finger-like formation as a consequence of Rayleigh-Taylor instability arising at its rim due to the inertia overcoming the stabilizing force of surface tension \citep{Yarin2006,Rioboo2001,Josserand2016} ultimately leading to its fragmentation as smaller droplets. On frigid surfaces as the drop equilibrates to the contact temperature the solidified layer additionally contracts which in the absence of pinning leads to bending \citep{deRuiter2018} as opposed to fracture when there is adhesion \citep{Ghabache2016}. Therefore, inertia of the spreading drop encounters additional resistance due to bending of the solidified layer, besides surface tension and fragmentation is delayed. 

For modeling such a scenario, we begin by approximating the volume of the spreading drop as a cylindrical pancake with diameter $D$ and height $h$ where, $D_{max}/2$, $D_{max}/3$ and $2V_0$ are chosen as the approximate average length scales (for $D$ and $h$) and horizontal spreading velocity scale ($V$) from the moment inertia begins to compete with solidification and surface tension just upon impact (at $t = 0$) until fingers begin to form (at $t = t_{max}$). To obtain these scales, we consider time averaged values from $t = 0$ to $t_{max}$. Thus, $D= D_{max}/2$ (average of $D=0$ at $t=0$ and $D=D_{max}$ at $t= t_{max}$) and $V = 4V_0/2$ (average of $V=4V_0$ at $t=0$ and $V << V_0$ at $t = t_{max}$) where, the value $4V_0$ is motivated by the fact that the drop upon impact experiences an initial acceleration horizontally such that $V$ is related to the vertical impact velocity as, $|V_0 \textrm{cot}\;(\theta_a)|$ \citep{Roux2004} with the apparent contact angle $\theta_a$ being approximately $166^{\textrm{o}}$ for our (and superhydrophobic) surfaces. 

Our experiments indicate that near fragmentation-bounce transition at $t = t_{max}$, $h \approx D_0/3$, and related to $D_{max}$ as $D_{max}/D_0 \approx 2.5$ (more precisely, for water and decane this ratio is 2.1 and 2.8 respectively), from where we get the time-averaged pancake thickness as $h \approx (D_0 + D_0/3)/2 = D_{max}/3$ which is the more appropriate form compared to the one ($h_{pan}$) used later using mass conservation as the drop flattens more and develops corrugation on the rim. This implies that the kinetic energy is given by, $(1/2)(\rho_l )(\frac{\pi}{4} \frac{D_{max}^2}{4} \frac{D_{max}}{3})(4V_0^2)$ which, for fragmentation to occur needs to exceed the sum of, (\textit{i}) elastic bending energy stored in the solidified layer of average thickness $\delta$ (in m), $Y \delta^3/12\left(1 - \gamma^2\right)$ where, $Y$ is the Young's modulus (in N$\cdot$m\textsuperscript{-2}) and  $\gamma$ is Poisson's ratio as well as (\textit{ii}) the surface energy of the cylinder, $\sigma_l (\frac{\pi}{4}) (\frac{D_{max}^2}{4})$ finally resulting in the following criterion for fragmentation (scaled by $D_0^3$),
\begin{equation} \label{eqn:1}
\dfrac{D_{max}^3}{D_0^3} > \dfrac{2Y\delta^3}{\pi\rho_l V_0^2D_0^3\left(1 - \gamma^2\right)} + \dfrac{3\sigma_l D_{max}^2}{2\rho_l V_0^2 D_0^3}
\end{equation}
Note that the term (\textit{i}), elastic bending energy signifies the energy associated with the \textit{unbending} of the solidified layer of the lower portion of area $A$ of the curved drop (dark blue shaded region in schematic Fig. \ref{Fig8}\textcolor{blue}{(\textit{ii})}) and is also multiplied by $D_{max}^2\kappa^{2} \approx 1$ as, $\kappa \sim D_{max}^{-1}$ and area, $A \sim D_{max}^2$ shown in the schematic Fig. \ref{Fig8}\textcolor{blue}{(\textit{iii})}.
\begin{figure} [htbp!]
\centering
\includegraphics[width = \textwidth]{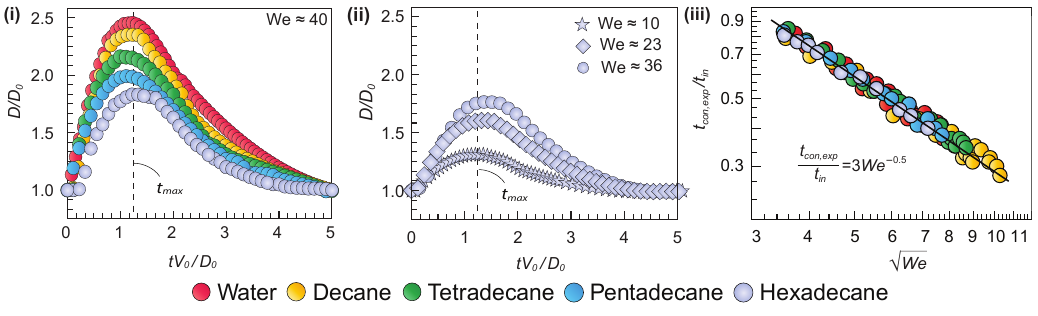}
\caption{\label{Fig7} Variation of dimensionless spreading diameter, $D/D_0$ with dimensionless time, $tV_0/D_0$ for drop bouncing, (\textit{i}) at constant $We \approx 40$, different liquids and, (\textit{ii}) for hexadecane and different $We$, the dotted vertical line at $t = t_{max}$ denotes the time at maximum spread, $D = D_{max}$  (\textit{iii}) ratio of the experimental contact time, $t_{con, exp}$ with the inertio-capillary time, $t_{in}$ showing its scaling with $\sqrt{We}$.}
\vspace{-3mm}
\end{figure} 

Solving the above equation requires determining the temporal evolution of solidification, $\delta\left(t\right)$ as the drop spreads from $0	\rightarrow D_{max} \rightarrow D_0$ and evaluating it at the time, $t_{max}$ corresponding to $D\approx D_{max}$ which is when fingers begin to appear (see Fig. \ref{Fig3} \textcolor{blue}{(\textit{iii})}). 

To estimate $t_{max}$, we analyzed the temporal evolution of the spreading droplet diameter ($D$) variation with time ($t$) for all the liquids. Examples of this analysis in the form of a non dimensional  drop diameter ($D/D_0$) as a function of non-dimensional time ($tV_0/D_0$) for all liquids at the same $We$, and for the same liquid (hexadecane) at different $We$ are shown in Figs. \ref{Fig7}\textcolor{blue}{(\textit{i})} and \textcolor{blue}{(\textit{ii})} respectively. For all the liquids, we find that the total contact time of drops, $t_{con} = 5D_0/V_0$ and $t_{max} = 5D_0/4V_0$ representing $1/4^{th}$ of its oscillation \citep{Richard2000}. Further analysis of the ratio of the experimental contact time, $t_{\textrm{\textit{con,exp}}}$ with the inertio-capillary time scale, $t_{in} (= \sqrt{\rho_l D_0^3/\sigma_l})$ as a function of $\sqrt{We}$ showed that $t_{\textrm{\textit{con,exp}}}$ scales as $We^{-1/2}$ (see Fig. \ref{Fig7} \textcolor{blue}{(\textit{iii})}). Using simple algebra this scaling simplifies to $t_{con, exp} \sim D_0/V_0$ (replaced by $t_{con}$ hereon), supporting the choice of this time scale in subsequent analysis.

We emphasize that the choice of $t_{con}$ scaling as $D_0/V_0$ is neither counter-intuitive nor does it contradict existing works. Impact of dry ice which have been likened to impacts on superhydrophobic surfaces have expressed the rebound (or contact) time as $\sqrt{\rho_lD_0^3/\sigma_l}$ \citep{Clanet2004} which seems to be odds with $D_0/V_0$. On closer inspection, we see that both are in essence the same with $D_0/V_0$ being the more generalized form. To understand this in detail we note that the inertio-capillary velocity scale, $V_0$ is given by $\sqrt{\sigma_l/\rho_lD_0}$ for a length scale given by the drop diameter, $D_0$ which means that $D_0/V_0$ produces $\sqrt{\rho_lD_0^3/\sigma_l}$ as the time inertio-capillary time scale, $t_{in}$. A natural query that arises from this explanation is: why we do not use $t_{in}$ for our analysis, consistent with literature. The reason for it is that solidification makes rebound calculations more complicated. Surface energy required to achieve rebound needs to not only equal the diminished inertia due to solidification but also overcome bending stresses (see Section \ref{Sec6} for details). In earlier works\citep{Antonini2013}, effects of solidification in rebound from dry ice have been largely unexplored as the choice of liquids (for instance, water) used for experimentation experience minimal solidification upon impact and therefore obviating the need to account for dissipation arising due to it. That being said, research on rebound of viscous drops does exist \citep{Jha2020}, and a contact time, $t_{con} = \left(1+\tfrac{1}{8}Oh^2\right)t_{in}$ which corrects $t_{in}$ by the prefactor, $\left(1+\tfrac{1}{8}Oh^2\right)$ for small $Oh$ to accommodate effects of viscous dissipation. In our work, similar correction could be made to $t_{in}$ to account for dissipation due to solidification however since the scale, $D_0/V_0$ absorbs all these effects we choose that as the contact time. In a future study, an explicit relation between $t_{con}$ and $t_{in}$ for non-isothermal drops impacts in general, could be explored and derived.

Returning to deriving our criterion for fragmentation we invoke Stefan's condition \citep{Mills2016} to evaluate $\delta$ at $t_{max} = 5D_0/4V_0$, wherein the latent heat of fusion equates to the conduction heat transfer across the solidified layer with a temperature difference, $\Delta T = T_{di} - T_m$ where, $T_{di}$ is the substrate temperature. From Eq. \ref{eqn:delfin} this results in $\delta = 2\lambda \lambda_c \sqrt{\alpha_s t}$  where, $\lambda (= \sqrt{Ste/2})$ accounts for the contact resistance limited solidification, $\lambda_c (= 0.2We^{2/5})$ is a constant multiplier to the prefactor $\lambda$. Using the defintion of \textit{Ste} and \textit{Pe} we obtain an expression for non-dimensional solidified thickness, $\overline{\delta} = \delta/ D_0 = \lambda_c\sqrt{5 Ste/2Pe}$ where, $\lambda$ is replaced by $\sqrt{Ste/2}$. Physically, higher $Ste$ implies faster solidification and higher $Pe$ indicates a smaller depth to which the effects of the substrate temperature penetrate. Although low velocity (i.e. smaller $Pe$) would mean a thicker thermal boundary layer, the contact resistance of the drop/dry ice interface (accounted by $\lambda_c$) limits its growth. The overall consequence of this competition is an increase in $\delta \sim V_0^{4/5}$resulting from, $\overline{\delta} = 0.32\;We^{2/5}\sqrt{Ste/Pe}$ which includes correction for contact resistance of the drop/dry ice interface as shown in Section \ref{Sec4}.

Experiments show that the transition from bouncing to fragmentation \textit{We} for water and decane are approximately 30 and 70 while the ratio of $D_{max}/D_0$  are 2.1 and 2.8 respectively (see Section \ref{Sec3}, Fig. \ref{Fig3} \textcolor{blue}{(\textit{v})} and \textcolor{blue}{(\textit{vi})}). This means the corresponding term, $(D_{max}/D_0)^3$ is approximately 10 and 22 for water and decane. At $t = t_{max} = t_{con}/4 \approx 5D_0/4V_0$ (see Fig. \ref{Fig8}\textcolor{blue}{(\textit{i}) - (\textit{iii})}), $\delta$ = 15 $\mu$m (water), 50 $\mu$m (decane) for a drop of $D_0$ = 2.3 mm impacting at $V_0$ = 1 m$\cdot$s\textsuperscript{-1}. Using $Y$ = 9 (ice), 0.2 (paraffin wax for solid decane) $\times$ 10\textsuperscript{9} Pa and $\gamma$ = 0.5, the term $2Y\delta^3/\pi\rho_l V_0^2D_0^3\left(1 - \gamma^2\right)$ evaluates to 10 for water and 16 for decane. On the other hand, at the transition \textit{We}, we find that the term $3\sigma_l D_{max}^2/2\rho_l V_0^2 D_0^3$ amounts to 0.22 for water and 0.16 for decane. From these we conclude that the second term, $3\sigma_l D_{max}^2/2\rho_l V_0^2 D_0^3$ of Eq. \ref{eqn:1} can be dropped for our impact conditions close to fragmentation. Consequently, inequality 1 transforms to, 
\begin{equation} \label{eqn:2}
\dfrac{D_{max}}{D_0} > \chi_1 \left(\lambda_c\sqrt{\dfrac{Ste}{Pe}}\right)
\end{equation}
where, $\chi_1 = \textcolor{black}{\sqrt{5/2}} \left[2Y/\pi \left(1 - \gamma^2\right) \rho_l V_0^2\right]^{1/3}$. Inequality \ref{eqn:2} suggests that as the thickness of the solidified layer increases, fragmentation of the drop is delayed, in agreement with our experimental observations for water and decane (see Fig. \ref{Fig8} \textcolor{blue}{(\textit{iv})}). 

Note that our criterion excludes splashing considerations based on \textit{lifting} of the lamella and a difference between the tip speed and the rate of increase of the wetted area \citep{Riboux2014, Garcia2021}. Solidification of the liquid layer close to the supercooled surface forestalls detachment of the lamella as its tip speed reduces drastically \citep{Kant2020a, Kant2020b} thereby preventing its lifting. Details regarding this are also described in the introduction, \ref{Sec1}. So Eq. \ref{eqn:2} predicts a simpler criterion for fragmentation. \textcolor{black}{Further note that, other interfacial energies like that of the solidified liquid with the dry ice substrate and solidified liquid with the unsolidified liquid do not play a part in our considerations for transition to fragmentation as they are balanced internally by the latent heat which equals the free surface energy required to create these new interfaces. Additionally, axial strain in the solidified splat is minimal and only the energy required to bend it is significant which is provided by the liquid surface tension ($\sigma_l$). Also, surface energy of the solid ($\sigma_s$) is insignificant due to smaller curved surface area. Therefore, in the spreading process, only changes in surface energy of the liquid manifests itself as a major factor along with elastic bending energy of the solidified layer which is balanced with the kinetic energy.}
\begin{figure} [t!]
\centering
\includegraphics[scale=0.75]{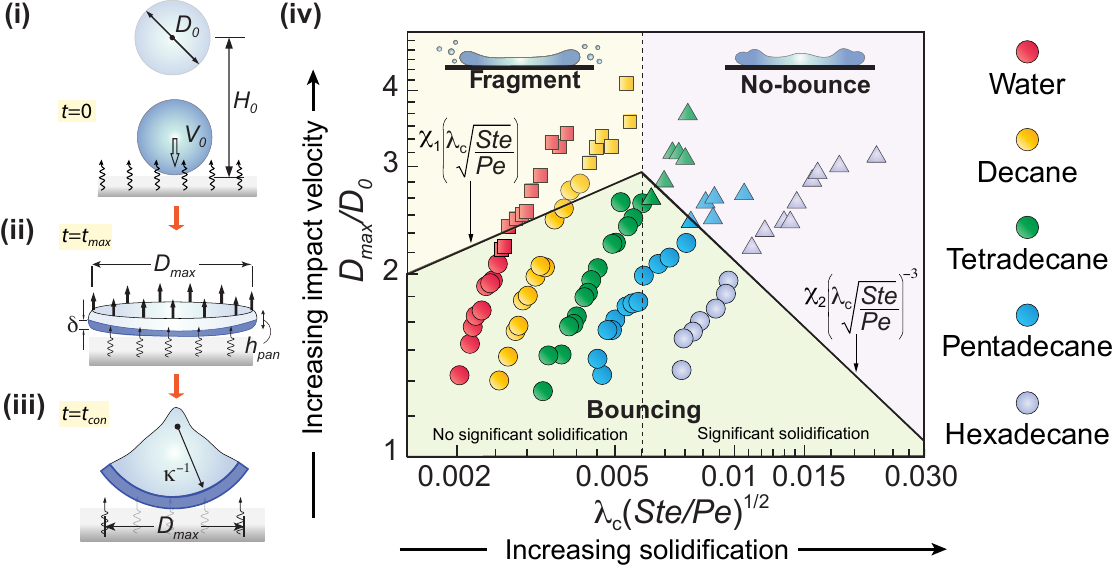}
\caption{\label{Fig8} (\textit{i}) Drop just before and after impact from height, $H_0$, with initial diameter, $D_0$ and, impact velocity, $V_0$ (\textit{ii}) Sketch of the flattened drop of total thickness, $h_{pan}$ showing the surface tension force acting on it (\textit{iii}) The final shape of the solidified layer of thickness, $\delta$ (with radius of curvature $\kappa^{-1}$) before rebound where $\delta << h_{pan}$. (\textit{iv}) Regime map showing three distinct impact outcomes affected by partial solidification - (\textit{a}) Fragmentation (in light yellow) (\textit{b}) No-bounce (in light purple) (\textit{c}) Bouncing with solidification (in light green) . The dotted vertical line represents the region beyond which the solidified layer thickness, $\delta$, is $\approx$ 14 $\mu m$ and bold lines correspond to the regime boundaries for fragmentation and no-bounce given by the inequalities \ref{eqn:2} and \ref{eqn:d_max_ste_pe}.}
\vspace{-3mm}
\end{figure} 
\subsection{Transition boundary between rebound and no-bounce}\label{Sec5b}
Delayed fragmentation implies bouncing is observed over a larger range of impact velocities or $We$ when solidification is limited but significant. But, an increase in \textit{We} leads tetradecane, pentadecane and hexadecane drops to transition from \textit{bounce} to a \textit{no-bounce} state without undergoing fragmentation (see Fig. \ref{Fig2}\textcolor{blue}{(\textit{ii})}, SM Movie 3 for hexadecane \bibnotemark[1] where this is pronounced). We suggest that as solidification increases, the energy required to deform, bend and wrap the solidified layer around the drop also increases such that the drop instead of bouncing (see Fig. \ref{Fig2}\textcolor{blue}{(\textit{ii})}) rests on the surface (see Fig. \ref{Fig2}\textcolor{blue}{(\textit{iv})}) without pinning. 

To determine the conditions under which \textit{no-bounce} would occur, we consider the flattening out of the drop into a pancake (see Fig. \ref{Fig8}\textcolor{blue}{(\textit{ii})}) with diameter, $D_{max}$ and thickness $h_{pan} \approx  D_0^3/D_{max}^2$, obtained by conserving mass before and after impact. The mass of the solidified layer ($\delta$) is a small fraction of the total mass and hence ignored. Bouncing with drop solidification would require the available bending capillary torque due to surface tension, $\textcolor{black}{\pi} \sigma_l D_{max} h_{pan}$ \citep{Bico2018, Chen2010} to overcome the flexural rigidity which is mathematically same as the elastic bending energy, $Y \delta^3/12\left(1 - \gamma^2\right)$. The solidified layer thickness, $\overline{\delta} \textcolor{black}{(= \lambda_c\sqrt{10Ste/Pe})}$ is determined at the drop contact time, $t_{con} \approx 5D_0/V_0$ demonstrated in Fig. \ref{Fig8}\textcolor{blue}{(\textit{i})}, \textcolor{blue}{(\textit{ii})} and, \textcolor{blue}{(\textit{iii})}.
and used in the remainder of this paper where we are concerned with rebound. In non-dimensional terms, the above arguments lead to,
\begin{equation} \label{eqn:d_max_ste_pe}
\dfrac{D_{max}}{D_0} < \chi_2 \left(\lambda_c\sqrt{\dfrac{Ste}{Pe}}\right)^{-3}
\end{equation}
where, $\chi_2 = \frac{3\pi\sqrt{10}}{25}\sigma_l(1-\gamma^2)/Y D_0$. Plotting Eq. \ref{eqn:d_max_ste_pe} in Fig. \ref{Fig8} \textcolor{blue}{(\textit{iv})} we see that it predicts the transition from no-bounce to bounce well. \textcolor{black}{In development of the above expression contribution of surface tension of the solidified liquid and interfacial tension of the solidified liquid with the unsolidified liquid are contingent to their dominant contributions in creating a torque to bend the solidified layer. Due to the small thickness of the solidified layer, they act at shallow angles with a small vertical component and therefore do not create a substantial torque compared to the liquid surface tension which acts dominantly around the periphery of the spreading drop. The lines of surface tension of the solidified liquid drop and the dry ice substrate pass through the center of the axisymmetric splat about which bending is considered. This creates no torque as the moment arm is zero thereby allowing us to neglect it.}

\section{Drop spreading}\label{Sec6}
Solidification affects both spreading and rebound of drops which can be quantified using - maximum spread ($D_{max}$) of the drop on impact \citep{Clanet2004, Antonini2016} and, its rebound height ($H_1$). To determine the arrest in the spread of the diameter of the deforming drop upon impact, we consider the Pad\'{e} approximant that interpolates between limits where a drop bounces in absence of solidification and the other when its completely solidified yet capable of bouncing back.
\begin{figure}[htp!]
   \centering
   \vspace{0pt}
    \includegraphics[scale=1.0]{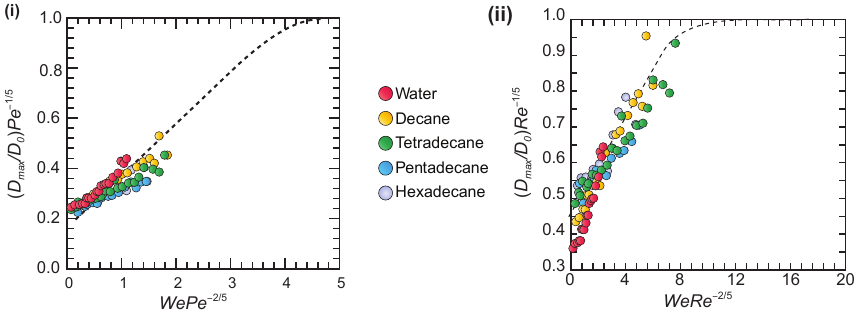}
   \vspace{0pt}
  \caption{ \label{Fig9}Scaling using (\textit{i}) $Pe^{1/5}$ and $We^{1/2}$, (\textit{ii}) $Re^{1/5}$ and $We^{1/2}$ as the asymptotic limits for Pad{\'e} approximant interpolation (dotted lines).}
  \vspace{-3mm}
\end{figure}

While, $We^{1/2}$ is surely the limit for bouncing with no solidification the other limit can have potentially other options. To demonstrate the success of scaling choosing $We^{1/2}$ and $WePe^{1/2}Ste^{-1/2}$ as the asymptotic limits, we consider other asymptotic limits and show their relatively poor performance. The unsatisfactory scaling of data using $Pe^{1/5}\;\textrm{or}\;Re^{1/5}$ as one of the asymptotic limits besides $We^{1/2}$ is tested here. Before we proceed it is important to note that the role of surface tension in our scaling model for drop spreading is significant and the scaling relations, $D_{max}/D_0 \sim Pe^{1/5}\;\textrm{or}\;Re^{1/5}$ only consider solidification and sticking \citep{Thievenaz2019, Gielen2020} however we also test them as potential options to confirm the correct scaling as we show later in this section.

(\textit{a}) \textbf{Interpolation between $We^{1/2}$ and $Pe^{1/5}$}: The rationale behind the scaling for $D_{max}/D_0 \sim We^{1/2}$ \citep{Clanet2004} and $D_{max}/D_0 \sim Pe^{1/5}$ \citep{Gielen2020} is documented in literature. Fig. \ref{Fig9}\textcolor{blue}{(\textit{i})} shows this scaling which has an $R^2 \approx 0.79$ and poor compared to the $R^2 \left( \approx 0.97\right)$ obtained for the interpolation using $We^{1/2}$ and $WePe^{1/2}Ste^{-1/2}$ as the asymptotic limits.

(\textit{b}) \textbf{Interpolation between $We^{1/2}$ and $Re^{1/5}$}: Another candidate when the drop is solidified and flattened is $Re^{1/5}$ \citep{Clanet2004}. Although it does not explicitly include thermal effects it can still be viable candidate. Fig. \ref{Fig9}\textcolor{blue}{(\textit{ii})} shows this scaling which has an $R^2 \approx 0.83$ and poor compared to the $R^2 \left( \approx 0.97\right)$ obtained for the interpolation using $We^{1/2}$ and $WePe^{1/2}Ste^{-1/2}$ as the asymptotic limits.

Since we only consider conditions where complete drop rebound is observed, a more restrictive limit which continues to account for surface tension is more successful and is derived below. This is also confirmed by the fact that $\alpha_s Pe/\alpha_l Ste > We^{5/2}$ for our test conditions, which follows from the fact that the capillary time scale is indeed less than the time scale for arrest of the spread of the drop \citep{Gielen2020}. 

\begin{figure}[htbp!]
\centering
\includegraphics[scale=1.0]{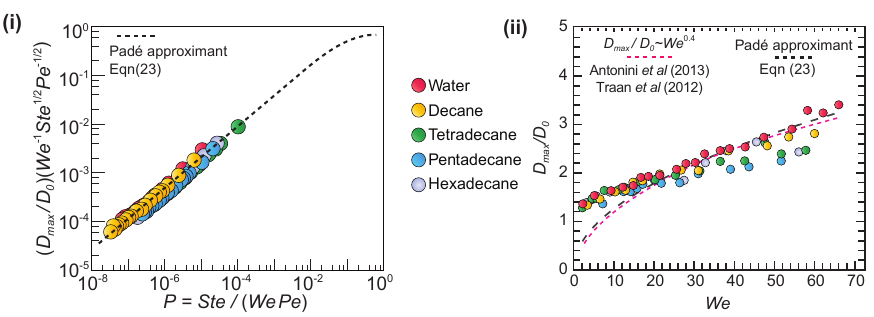}
\caption{\label{Fig10} Scaling for maximum non-dimensional spread, $D_{max}/D_0$ using Pad\'{e} approximant \citep{Laan2014} and given by (dotted line), Eq. \ref{EqPade}. The Pad\'{e} approximant closely follows the scaling relation, $D_{max}/D_0  \sim We^{0.4}$ \citep{Antonini2013, Tran2012}}
\vspace{-2mm}
\end{figure}

Therefore, in scaling $D_{max}/D_0$ shown in Fig. \ref{Fig10} we choose $We^{1/2}$ and $WePe^{1/2}Ste^{-1/2}$ as the asymptotic limits for Pad{\'e}'s approximant which represent no solidification ($We^{1/2}$) and complete solidification ($WePe^{1/2}Ste^{-1/2}$). The first limit given by $D_{max}/D_0 \sim We^{1/2}$ is obtained from the complete conversion of kinetic energy into surface energy and shown \citep{Lee2016, Laan2014} to be more applicable than $D_{max}/D_0 \sim We^{1/4}$ \citep{Clanet2004}. 
For the second limit, we tested the possibility of $D_{max}/D_0 \sim Re^{1/5}$ and $D_{max}/D_0 \sim Pe^{1/5}$ as possible options which have been traditionally used for completely solidified splats \citep{Thievenaz2019, Gielen2020} however both yield unsatisfactory scaling with $R^2 \leq 0.85$ and do not depict a situation where the drop rebounds. Consequently, we consider a hypothetical situation where the drop solidifies almost instantly as it starts to spread on a supercooled surface such that $D_{max} \approx D_0$ and $h_{pan} \approx \delta$. In this limit, the energy balance between total energy of the drop before and after contact reads as, $\rho_l D_0^3 V_0^2 + \sigma_l D_0^2 \sim \sigma_s D_{max}^2 + \sigma_s h_{pan} D_{max}$, where $\sigma_s$ (in N$\cdot$m\textsuperscript{-1}) is the surface tension of the test liquids in their solidified state. Ignoring the slight differences between $\sigma_s$ and $\sigma_l$ (see \bibnotemark[1] Section S1) it follows that $\sigma_l D_0^2 \approx \sigma_s D_{max}^2$ which results in the energy balance, $\rho_l D_0^3V_0^2 \sim \sigma_s \delta D_{max}$ that can be recast as, $D_{max}/D_0 \sim We ({Pe/Ste})^{1/2}$. 

The interpolating Pad{\'e} approximant function corresponding to these limits is given by,
\begin{equation}\label{EqPade}
\dfrac{D_{max}}{D_0} We^{-1} ({Pe/Ste})^{-1/2} = \dfrac{P^{1/2}}{0.0025 + P^{1/2}} \textrm{  where, } P = We^{-1}Pe^{-1}Ste
\end{equation}
(see Fig. \ref{Fig10}\textcolor{blue}{(\textit{i})}) and corresponds to an $R^2 \approx \textrm{0.98}$. The slight preponderance of data to the left axis of the plot shows more cases with partial solidification and bouncing. Note that, comparing our interpolation with previous studies on drop impact and rebound on a gaseous cushion in non-isothermal conditions \citep{Antonini2013, Tran2012} we see that the scaling for the spread factor, $D_{max}/D_0 \sim We^{2/5}$ stated therein has a close overlap with our Pad\'{e} fit (see Fig. \ref{Fig10}\textcolor{blue}{(\textit{ii})}). 
 
\section{Drop Rebound}\label{Sec7}
\begin{figure}[htp!]
\vspace{-1mm}
\centering
\includegraphics[scale=1.2]{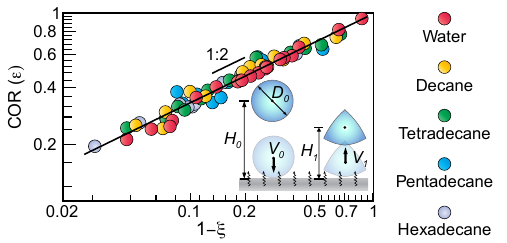}
\caption{\label{Fig11} Comparison of experimental data and theoretical prediction for coefficient of restitution ($\epsilon$) for the first bounce of a partially solidified drop.}
\vspace{-3mm}
\end{figure}

In the last part of our discussion we investigate the role of drop solidification (in terms of $Ste$) on the rebound height (in terms of $H_1$) as shown in Fig. \ref{Fig4} \textcolor{blue}{(\textit{i})} and \textcolor{blue}{(\textit{ii})} (also see SM Movie 2 \bibnotemark[1]). As discussed before, the thickness of the solidified layer, $\delta$ is related to $Ste$ and the impact velocity through $Pe$ or $We$ which motivates these investigations. To quantitatively understand this decrease in $H_1$ we develop a theoretical expression for the coefficient of restitution, COR ($\epsilon$) which includes the effects of solidification. COR is represented by the ratio of the rebound ($V_1$) and initial velocity ($V_0$) (see inset of Fig. \ref{Fig11}) and is indicative of the repellency of a surface such that superhydrophobic surfaces with low contact angle hysteresis demonstrate the highest $\epsilon \approx 1$ \citep{Richard2000}. 

To evaluate $\epsilon$ we consider consider kinetic energy after impact/rebound, $E_r (= mV_1^2/2)$ and the initial kinetic energy, $E_{in}(= mV_0^2/2)$ and the energy loss during rebound, $\Delta E$, which is the sum of contributions from loss of kinetic energy due to arrest of movement of the solidified layer, $(1/2)(\rho_s \pi D_{s}^2 \delta/4) V_0^2$ (where, $D_s \approx D_{max}/2$ is the average contact diameter as the drop spreads from 0 to $D_{max}$ and $\rho_s$ is the density of the solidified layer, $\delta$) and energy expended in bending the solidified layer, $Y \delta^3/12\left(1 - \gamma^2\right)$.With these considerations, $E_r$ can be simply written as $E_{in} - \Delta E$, which leads to, $\epsilon^2 = {1 - a_1\overline{\delta} - a_2\overline{\delta}^3}$ where, $a_1 = (3/8)\left(\rho_s/\rho_l\right)\left(D_{max}/D_0\right)^2 $ and $a_2 = Y/\pi \rho_l V_0^2\left(1 - \gamma^2\right)$. Denoting, $\xi = a_1\overline{\delta} + a_2 \overline{\delta}^3$ we may write concisely, $\epsilon = \sqrt{1 - \xi}$. Comparing our experimental results with theoretical predictions in Fig. \ref{Fig11} evaluated at $t = t_{con}$ we find that $1:2$ scaling between $\epsilon$ and $1- \xi$ is recovered

\section{Summary and Conclusions}\label{Sec8}

In summary, using a unique combination of ultra-low adhesive dry ice surface and alkanes as liquids, we isolated the role of solidification during drop impact. We further demonstrated that solidification within a drop, even though partial, dissipates its initial kinetic energy to delay fragmentation, reduces its spread, and decreases its coefficient of restitution – even suppressing the rebound entirely, thus providing a strategy to control the drop deposition by locally changing the substrate temperature in applications such as paint spraying and additive manufacturing. The fragmentation/no-bounce criterion and the maximum spreading diameter developed herein along with calculations on contact resistance limited depth of solidification are expected to be applicable to any liquid contact with supercooled substrates and especially those with ultra-low adhesion. They also serve to provide new insights into dissipation mechanisms in drop impact on supercooled, non-wetting surfaces. The findings in this work expand our current understanding which has limited itself to studies of drop rebound on dry ice to other previously unknown or unexplained outcomes such as fragmentation and no-bounce. This makes dry ice a versatile platform which can lead to a whole gamut of scenarios unlike traditional supercooled surfaces which are restricted to pinning mediated adhesion. \textcolor{black}{The finding that despite partial solidification, drops can rebound from a surface on ultra-low adhesive surfaces can have significant implications for designing robust icephobic/anti-icing surfaces.} Lastly, we expect our results have a tremendous bearing on developing strategies to repel or enhance adhesion of a wide range of liquids undergoing liquid-solid phase change to the surfaces, including applications such as wax deposition, and liquid transport in microfluidic channels besides scenarios where fragmentation needs to be controlled.

\section*{Acknowledgment}

This work was supported by NSF (CAREER) award (no. 1847627) and Society in Science Branco Weiss Fellowship. Suhas Tamvada, Nikhil Shirdade and Navid Saneie contributed equally to this work. 
Additional data and material related to this paper can be requested from Varun Kulkarni (\href{varun14kul@gmail.com}{varun14kul@gmail.com}).

\bibliography{arXiv_Kulkarni_2024}
\end{document}


\maketitle
\thispagestyle{empty}


\section*{Supplementary Videos}
The videos for this work were recorded using Photron\textsuperscript{\textregistered} FASTCAM Mini AX high-speed camera at 1000 frames per second (fps) with a resolution of $1024 \times 1024$ pixels and exposure time of 5 $\mu s$. The time scale of the drop impact process was in excess of 10 $ms$ and a temporal resolution of 1 $ms$ was appropriate to capture the entire dynamics. A lens with infinite focus (InfiniProbe\textsuperscript{\textregistered} TS-160) and focal length ranging from infinity to 18 $mm$ and a magnification ranging from $0-16\times$ was attached to the camera such that 1 pixel $\approx$10 $\mu m$. The experiments were backlit using an LED (Nila-Zaila\textsuperscript{\textregistered}) illumination source which was diffused using diffuser plates. The videos/images thus captured were analyzed using an open source application, IMAGE J.

\begin{enumerate}[leftmargin=0\parindent]
	\item[] \textbf{\textcolor{blue}{Supplementary Video 1: Movie\_01.avi}}
	\vspace{2pt}
	\\
	Different outcomes of droplet impact on dry ice. Four distinct regimes can be seen (from left to right) (\textit{ii}) Bounce-back without appreciable solidification (Water, $We \approx 9$), $\delta < 20 \mu m$ (\textit{i}) Fragmentation (Decane, $We \approx 32$) (\textit{iii}) Bounce-back with appreciable solidification (Hexadecane, $We \approx 34$), $\delta > 20 \mu m$ (\textit{iv}) Sticking (Hexadecane, $We \approx 56$). 
	The video is played back at 10 fps.
		\vspace{2pt}
	\item[] \textbf{\textcolor{blue}{Supplementary Video 2: Movie\_02.avi}}
	\vspace{2pt}
	\\
	Different bounce height for different liquids, Water, Decane, Tetradecane, Pentadecane and Hexadecane at same $We \approx 20$ corresponding to a height of $2\;cm$ showing the effect of solidification. From left to right, the extent of solidification increases and leads to smaller rebound heights.
	The video is played back at 10 fps.   
	\vspace{2pt}
	\item[] \textbf{\textcolor{blue}{Supplementary Video 3: Movie\_03.avi}}
	\vspace{2pt}
	\\
	Drop deformation behavior for hexadecane corresponding to $We \approx $ 32, 56, 63, 78, 92 (from left to right) and decane corresponding to $We \approx $ 38, 54, 60, 80, 89 (from left to right). The video is played back at 5 fps. 
\end{enumerate}